\documentclass[
aps,
prl,
amsmath,
amssymb,
twocolumn,
floatfix,
superscriptaddress,
longbibliography,
nobibnotes
]{revtex4-1}

\pdfoutput=1

\usepackage{graphicx}
\usepackage{dcolumn}
\usepackage{bm}
\usepackage{epstopdf}
\usepackage{epsfig}
\usepackage{multirow}
\usepackage{color}
\usepackage[normalem]{ulem}
\usepackage{upgreek}
\usepackage{float}
\usepackage[]{subfig}
\usepackage[export]{adjustbox}
\usepackage{hyperref}

\hypersetup{
    colorlinks=true,
    linkcolor=black,
    citecolor=black,
    filecolor=black,
    urlcolor=black,
}

\begin{document}

\title{
Integrated multiplexed microwave readout of silicon quantum dots in a cryogenic CMOS chip
}

\author{A. Ruffino}
\thanks{Equally-contributing authors}
\affiliation{Advanced Quantum Architecture Laboratory, \'Ecole Polytechnique F\'ed\'erale de Lausanne, Rue de la Maladi\`ere 71b, 2002, Neuch\^atel, Switzerland}
\author{T.-Y. Yang}
\thanks{Equally-contributing authors}
\affiliation{Hitachi Cambridge Laboratory, Hitachi Europe Ltd., J. J. Thomson Avenue, CB3 0HE, Cambridge, United Kingdom}
\author{J. Michniewicz}
\affiliation{Cavendish Laboratory, University of Cambridge, J. J. Thomson Avenue, CB3 0HE, Cambridge, United Kingdom}
\author{Y. Peng}
\affiliation{Advanced Quantum Architecture Laboratory, \'Ecole Polytechnique F\'ed\'erale de Lausanne, Rue de la Maladi\`ere 71b, 2002, Neuch\^atel, Switzerland}
\author{E. Charbon}
\thanks{Equally-credited authors}
\affiliation{Advanced Quantum Architecture Laboratory, \'Ecole Polytechnique F\'ed\'erale de Lausanne, Rue de la Maladi\`ere 71b, 2002, Neuch\^atel, Switzerland}
\author{M. F. Gonzalez-Zalba}
\thanks{Equally-credited authors}
\affiliation{Hitachi Cambridge Laboratory, Hitachi Europe Ltd., J. J. Thomson Avenue, CB3 0HE, Cambridge, United Kingdom}
\affiliation{Quantum Motion Technologies, Nexus, Discovery Way, LS2 3AA, Leeds, United Kingdom}

\maketitle

\textbf{
Solid-state quantum computers require classical electronics to control and readout individual qubits and to enable fast classical data processing~\cite{arute2019nature,watson2018nature,reilly2015npjqi}. Integrating both subsystems at deep cryogenic temperatures~\cite{charbon2016iedm}, where solid-state quantum processors operate best, may solve some major scaling challenges, such as system size and input/output (I/O) data management~\cite{reilly2019iedm}. Spin qubits in silicon quantum dots (QDs) could be monolithically integrated with complementary metal-oxide-semiconductor (CMOS) electronics using very-large-scale integration (VLSI) and thus leveraging over wide manufacturing experience in the semiconductor industry~\cite{maurand2016natcomm}. However, experimental demonstrations of integration using industrial CMOS at mK temperatures are still in their infancy.
\newline\indent
Here we present a cryogenic integrated circuit (IC) fabricated using industrial CMOS technology that hosts three key ingredients of a silicon-based quantum processor: QD arrays (arranged here in a non-interacting 3$\times$3 configuration), digital electronics to minimize control lines using row-column addressing and analog LC resonators for multiplexed readout, all operating at 50 mK. With the microwave resonators (6-8 GHz range), we show dispersive readout of the charge state of the QDs and perform combined time- and frequency-domain multiplexing, enabling scalable readout while reducing the overall chip footprint.
\newline\indent
This modular architecture probes the limits towards the realization of a large-scale silicon quantum computer integrating quantum and classical electronics using industrial CMOS technology.
}

\begin{figure*}[t]
\centering
\includegraphics[width=\textwidth]{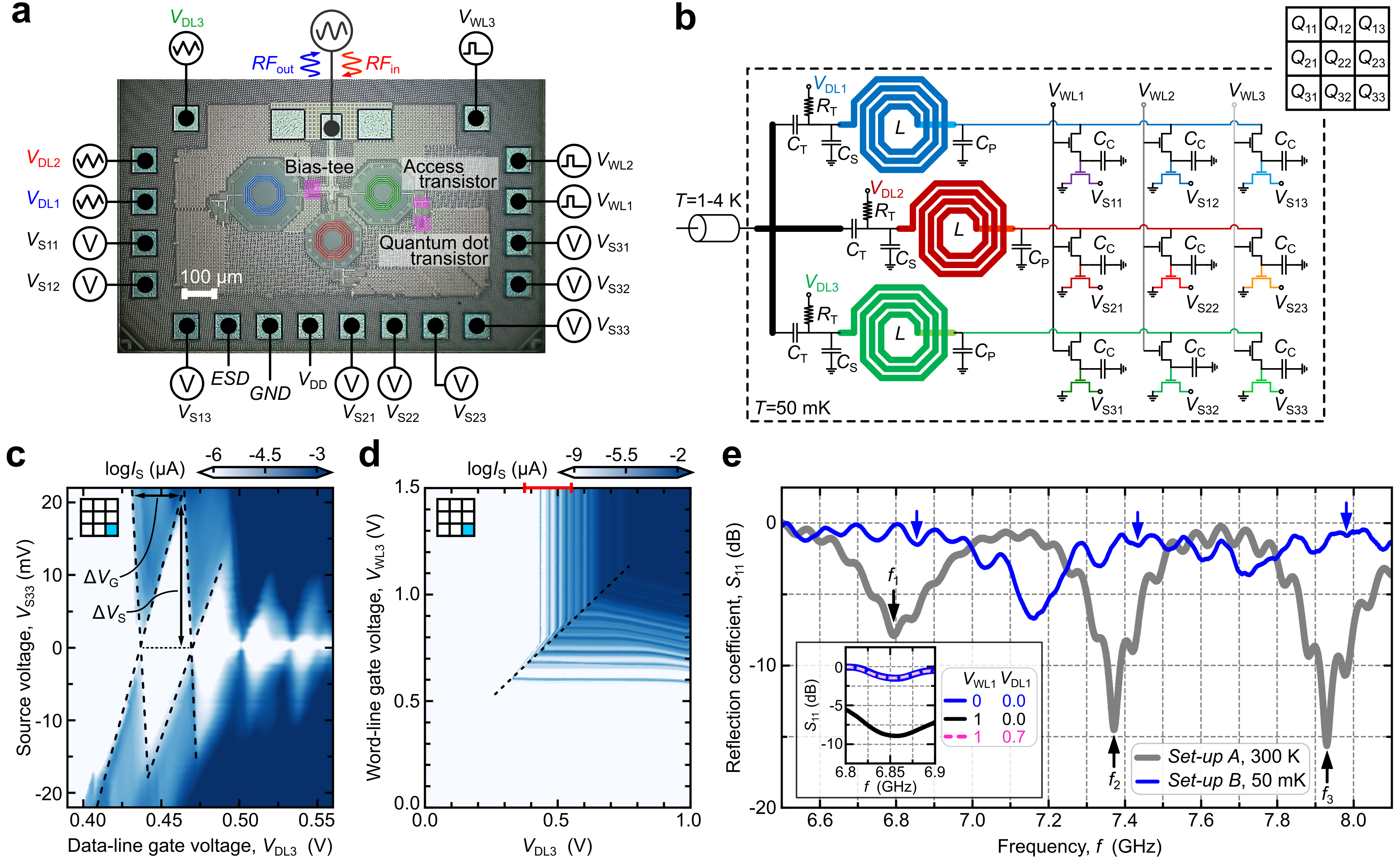}
\caption{\textbf{Fully-integrated cryogenic CMOS quantum-classical matrix.} \textbf{a,} Micrograph of the IC, with corresponding control signals. \textbf{b,} Schematic of the cryogenic CMOS readout matrix consisting of three integrated LC resonators, i.e., $Resonator_\mathrm{1}$ (blue), $Resonator_\mathrm{2}$ (red), and $Resonator_\mathrm{3}$ (green), access transistors $T_\mathrm{ij}$ and QD transistors $Q_\mathrm{ij}$, word-lines $V_\mathrm{WLj}$, data-lines $V_\mathrm{DLi}$, source voltages $V_\mathrm{Sij}$, bias-tees $R_\mathrm{T}$-$C_\mathrm{T}$, and storage capacitors $C_\mathrm{C}$. Inset: A 3$\times$3 grid representing the 3$\times$3 QD transistor array. \textbf{c,} Coulomb diamonds at $V_\mathrm{WL3}$=1.5~V with $V_\mathrm{DL3}$ ranging from 0.39 V to 0.56~V, as marked by the red line section in \textbf{d}. \textbf{d,} Stability diagram of $V_\mathrm{WL3}$ vs $V_\mathrm{DL3}$ with $V_\mathrm{S33}$=0.01~V. \textbf{e,} Frequency spectrum of the multi-resonator for \textit{Set-up~A} at 300~K and \textit{Set-up~B} at 50~mK, showing the 3 resonance frequencies. \textit{Set-up~A}: direct measurement with vector network analyzer (VNA) $\to$cable$\to$chip$\to$cable$\to$VNA. \textit{Set-up~B}: the chip is placed in the dilution fridge, and connected via VNA$\to$attenuator$\to$cable$\to$circulator$\to$chip$\to$circulator$\to$low noise amplifiers$\to$cable$\to$VNA (see Methods). The resonance frequencies are found to be ($f_{1}$, $f_{2}$, $f_{3}$)=(6.810, 7.374, 7.941)~GHz for ($Resonator_\mathrm{1}$, $Resonator_\mathrm{2}$, $Resonator_\mathrm{3}$) at 300~K (see Supplementary Information S2). The black and blue arrows indicate the resonance frequencies at 300~K and 50~mK, respectively. Inset: zoom-in for $Resonator_\mathrm{1}$ at different $V_\mathrm{WL1}$ and $V_\mathrm{DL1}$.}
\label{fig:matrix}
\end{figure*}

Quantum computing is poised to be an innovation driver of the decade, given its theoretically-demonstrated capability to solve certain computational problems more efficiently than classical computers~\cite{montanaro2016npjqi}. However, constructing the required quantum hardware is one of the greatest technological challenges for the scientific community. Single electron spins isolated in silicon QDs are one of the most promising solid-state systems to achieve that goal: recent demonstrations of long coherence times~\cite{veldhorst2014natnano}, high-fidelity spin readout~\cite{urdampilleta2019natnano}, and one- and two-qubit gates~\cite{yoneda2018natnano,huang2019nature,zajac2018science}, fulfill the basic requirements to build a quantum computer approaching fault-tolerant thresholds~\cite{fowler2012pra}. Until now, silicon QDs have been typically fabricated using custom processes~\cite{kawakami2014natnano,veldhorst2014natnano}, but recent results have revealed they can be manufactured at scale using industry-compatible~\cite{maurand2016natcomm} or even industry-standard processes~\cite{yang2020edl,bonen2019edl}. This allows to leverage the integration capabilities of the semiconductor industry to scale up.
\newline\indent
Researchers have already produced blueprints of large-scale quantum computers in silicon~\cite{veldhorst2017natcomm,vandersypen2017npjqi,li2018sa}. The proposals share common concepts: MOS-based QD arrays to host the qubits, digital electronics for I/O data management and analog classical electronics for control and readout. Full system integration will bring the benefits of reduced footprint, ease of signal synchronization, reduced latency and minimized inter-chip wiring. However, the ultimate level of possible integration is still uncertain, given the reduced cooling power availability at mK temperatures for typical classical electronics. Exploring the limits of integration is hence paramount to realize any fully-fledged solid-state quantum processor~\cite{leguevel2020isscc,opremcak2018science}.
\newline\indent
For silicon, one approach has been to operate qubits at elevated temperatures ($\sim$1.2~K)~\cite{yang2020nature,petit2020nature} to enable larger cooling power budgets for the classical electronics~\cite{xue2020arxiv}, but so far this has come at the cost of reduced fidelity and/or coherence times. Alternatively, ICs that operate at mK temperatures have been produced for control~\cite{pauka2019arxiv}, readout~\cite{hornibrook2014apl}, signal multiplexing~\cite{pauka2020pra,potocnik2020arxiv,paqueletwuetz2020npjqi}, or single devices have been used for time-multiplexed readout~\cite{schaal2019natel}, but these circuits have not been fully integrated with the quantum devices.
\newline\indent
Here we present an IC fabricated using industrial 40\nobreakdash-nm CMOS technology that enables scalable multiplexed microwave readout of a non-interacting array of silicon QDs, all operating at mK temperatures in the same monolithic chip. The QDs are hosted in the channel of minimum size transistors and placed in a 3$\times$3 array. Individual QDs can be randomly addressed via digital transistors using a row-column architecture to minimize the number of inputs. The readout is performed using gate-based microwave reflectometry, a technique readily compatible with industrial CMOS technology~\cite{west2019natnano}. For the first time, we combine time- and frequency-domain multiplexing in one demonstration to minimize circuit footprint, otherwise compromised by using one readout resonator per qubit, while preserving a degree of parallel readout, ideal for quantum error correction.

\section*{A fully-integrated DRAM-like readout matrix architecture}

The proposed quantum-classical readout interface has been designed and fabricated in an industrial 40-nm bulk CMOS technology. Fig.~\ref{fig:matrix}a and Fig.~\ref{fig:matrix}b present the chip micrograph and its schematic, respectively. The modular architecture consists of a 3$\times$3 array of 9 identical cells, arranged in a row-column random-access configuration similar to a dynamic random-access memory (DRAM). Each cell contains a silicon QD device, i.e. $Q_\mathrm{ij}$ for $i,j$=1,2,3, implemented as a MOS transistor. Furthermore, the gate of each quantum device is connected to the source of an access MOS transistor with a channel length $L$=40~nm and a channel width $W$=15~$\upmu$m, allowing conditional access, and to a storage capacitor (200~fF), to store the voltage at the gate of the quantum device (see Supplementary Information S1). The access transistors in column $j$ are controlled by the same word-line signal $V_\mathrm{WLj}$. The cells in row $i$ are connected to both a shared data-line signal $V_\mathrm{DLi}$, providing a bias voltage to control the QDs through an on-chip bias-tee, and a LC resonator $Resonator_\mathrm{i}$, to perform microwave reflectometry readout~\cite{colless2013prl,wallraff2004nature}. When applying data-line voltage $V_\mathrm{DLi}$ and word-line voltage $V_\mathrm{WLj}$, a quantum device $Q_\mathrm{ij}$ is accessed, and its charge state can be read out via $Resonator_\mathrm{i}$. We also included independent source voltages $V_\mathrm{Sij}$ to perform a full electrical characterization of the individual QD devices in the quantum regime.
\newline\indent
This architecture reduces the resources needed for control and readout of $N$ quantum devices from $N$ to just $2\sqrt{N}$ lines and $\sqrt{N}$ resonators, respectively, substantially improving upon the circuit complexity of current paradigms.

\section*{Quantum-classical circuits in standard CMOS}

The QD devices are designed as minimum size ($W/L$=120~nm/40~nm) nMOS transistors, to minimize channel volume. The devices have a low threshold voltage $V_\mathrm{th}$ and they are implemented as single gate planar NPN transistors in a deep n-well, to isolate them from substrate noise. When we cool these devices down to 50~mK and apply a positive $V_\mathrm{DLi}$ voltage ($\sim V_\mathrm{th}$) with high $V_\mathrm{WLj}$, single electrons are trapped in the small area under the gate at the oxide-silicon interface, i.e. few-electron QDs~\cite{zwanenburg2013rmp} are formed, which we observe at low $V_\mathrm{Sij}$, as shown in Fig.~\ref{fig:matrix}c.
\newline\indent
Here we measured the source-drain current $I_\mathrm{S}$ as a function of the $V_\mathrm{DLi}$ and $V_\mathrm{Sij}$ voltages for the QD transistor $Q_\mathrm{33}$ and we observe Coulomb diamonds. From this plot, a charging energy $E_\mathrm{c}$=$e$$\cdot$$\Delta V_\mathrm{S}$=21.42~meV ($e$ is the electron charge) can be extracted, which enables electron trapping even at 4.2~K as $E_\mathrm{c}$$\gg$$k_\mathrm{B}T$ ($k_\mathrm{B}$ and $T$ are the Boltzmann constant and temperature, respectively). At 50~mK, all 9 devices in the matrix exhibit Coulomb blockade oscillations.
\newline\indent
In Fig.~\ref{fig:matrix}d we demonstrate the functionality of the access transistor-QD cell and highlight the allowed and forbidden regions for readout. When $V_\mathrm{WL3}-V_\mathrm{DL3}\!<$0.277~V, the access transistor channel resistance $R_\mathrm{acc}$ becomes comparable to the QD transistor gate leakage resistance $R_\mathrm{G}$, therefore the effective gate voltage for the QD transistor $V_\mathrm{DL,eff}$=($V_\mathrm{DL}$$\cdot$$R_\mathrm{G}$)/($R_\mathrm{acc}$+$R_\mathrm{G}$) is reduced. So, the QD becomes effectively decoupled from the data-line (see Supplementary Information S1). However, when $V_\mathrm{WL3}-V_\mathrm{DL3}$ is far higher than this threshold, i.e., above the dashed black line in Fig.~\ref{fig:matrix}d, then $V_\mathrm{DL,eff}$$\approx$$V_\mathrm{DL}$, and Coulomb blockade oscillations are observed as a function of $V_\mathrm{DL3}$.
\newline\indent
This demonstrates the realization of a quantum-classical integrated circuit in industrial bulk CMOS technology at 50~mK.

\begin{figure}[t]
\centering
\includegraphics[width=\columnwidth]{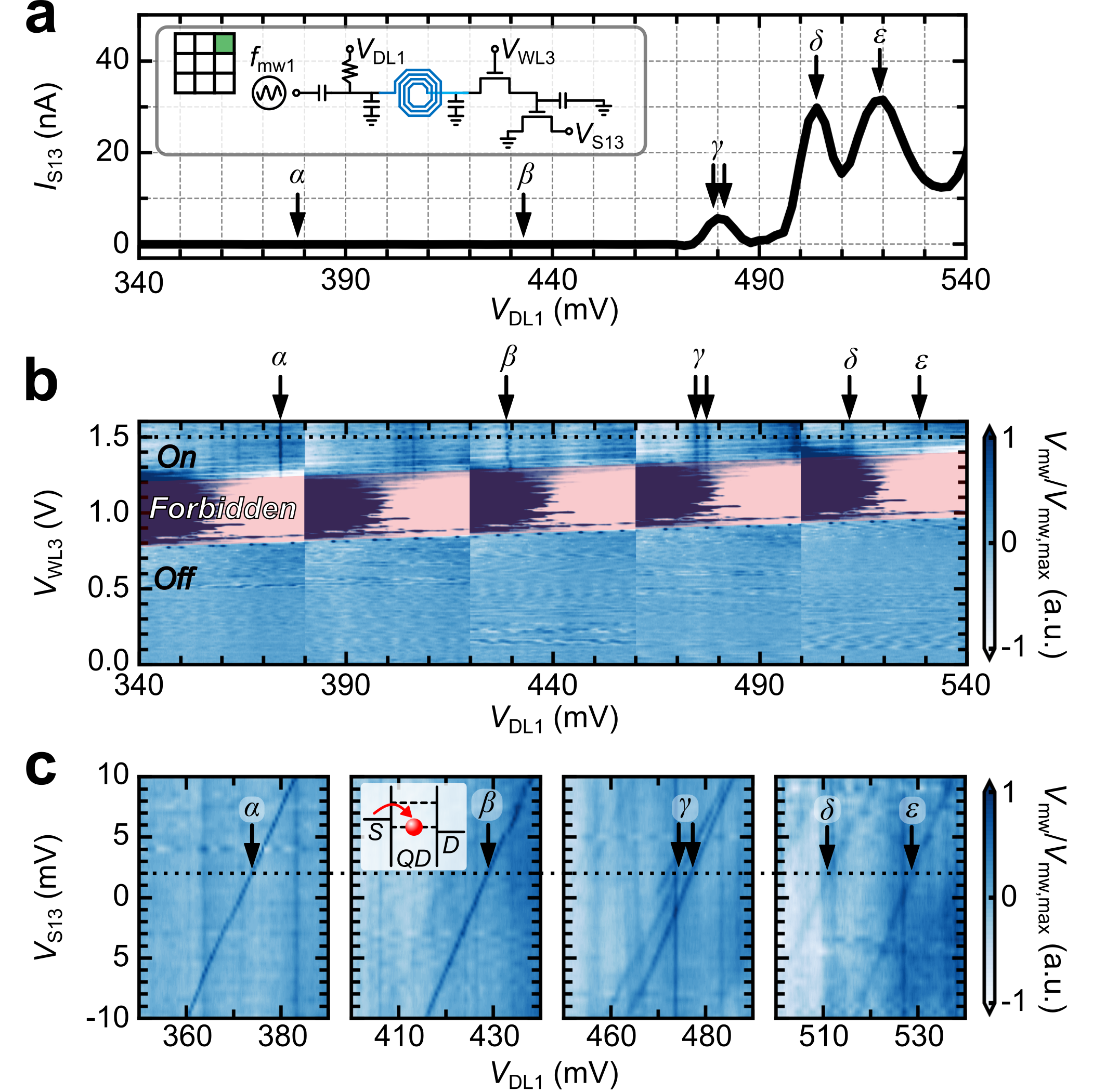}
\caption{\textbf{Integrated gate-based reflectometry.} \textbf{a,} DC transport measurement for device $Q_\mathrm{13}$: $I_\mathrm{S13}$ vs $V_\mathrm{DL1}$. \textbf{b,} Normalized microwave reflected voltage as a function of $V_\mathrm{WL3}$ and $V_\mathrm{DL1}$ at $V_\mathrm{S13}$=2~mV. This diagram is the union of five individual $V_\mathrm{WL3}$ vs $V_\mathrm{DL1}$ measurements, resulting in the repetitive blue/pink pattern in the \textit{Forbidden} region. \textbf{c,} Microwave reflectometry stability diagrams as a function of $V_\mathrm{S13}$ and $V_\mathrm{DL1}$ for the Coulomb peaks $\alpha$, $\beta$, $\gamma$, $\delta$, and $\varepsilon$ at $V_\mathrm{WL3}$=1.5~V. Inset: energy diagram of $Source$-$QD$ transition generating the signal $\beta$. The measurements are performed at 50~mK, and the carrier frequency is $f_\mathrm{mw1}$=6.877~GHz.}
\label{fig:reflectometry}
\end{figure}

\section*{Radio-frequency characterization}

We designed the resonators $Resonator_\mathrm{i}$ to have different resonant frequencies $f_\mathrm{i}$, so as to enable frequency selective readout over the corresponding rows. Fig.~\ref{fig:matrix}e shows the frequency spectrum for the integrated LC resonators. At 300~K, the reflection coefficient $S_\mathrm{11}$ presents 3 minima at the resonant frequencies ($f_{1}$, $f_{2}$, $f_{3}$)=(6.810, 7.374, 7.941)~GHz (see grey trace). At 50~mK, when all $V_\mathrm{WLj}$ are set to 0~V, the spectrum shows high mismatch (see blue trace), while, by turning on, for example, the column voltage $V_\mathrm{WL1}$, all the access transistors on that column turn on, modifying the total impedance of the system and hence the reflected power (see black trace in the inset). The resonators are designed to match the high impedance of the QD device gate to the 50~$\Omega$ microwave input when the access transistors are on, allowing maximum power transfer and enhanced sensitivity~\cite{ares2016pra}. To demonstrate the frequency selectivity of the readout, we then increase the data line voltage $V_\mathrm{DL1}$ to turn off the access transistor $T_{11}$ (see pink dashed line in the inset). Only when $T_{11}$ is switched off, the reflected power at $f_\mathrm{1}$ returns to its original value. At 50~mK, the probing frequencies are identified as ($f_{1}$, $f_{2}$, $f_{3}$)=(6.872, 7.420, 7.951)~GHz.
\newline\indent
Frequency-multiplexing interfaces have been previously reported \cite{hornibrook2014apl}, however typically readout frequencies below 1~GHz have been used \cite{west2019natnano}. More recently, silicon QDs were interfaced with microwave resonators in the 6-8~GHz range to explore coherent spin-photon interactions \cite{samkharadze2018science,mi2018nature}. Although these resonators enabled fast state readout~\cite{zheng2019natnano}, hybrid manufacturing was necessary. Here, the resonators and the QDs are co-integrated in the same industrial CMOS process.
\newline\indent
Operating at higher readout frequency has several advantages. Firstly, it reduces the footprint of the inductors, the largest elements of the architecture. Furthermore, the resonator quality factor, critical for the sensitivity of the technique~\cite{ahmed2018pra}, is higher for smaller inductors used at higher frequencies. Our quality factors are modest ($Q<100$), compared to superconductor-based resonators, but show the state-of-the-art of what can be achieved with standard CMOS. Finally, the sensitivity of gate-based dispersive readout is higher at higher frequencies \cite{gonzalez-zalba2015natcomm}.

\begin{figure}[t]
\centering
\includegraphics[width=\columnwidth]{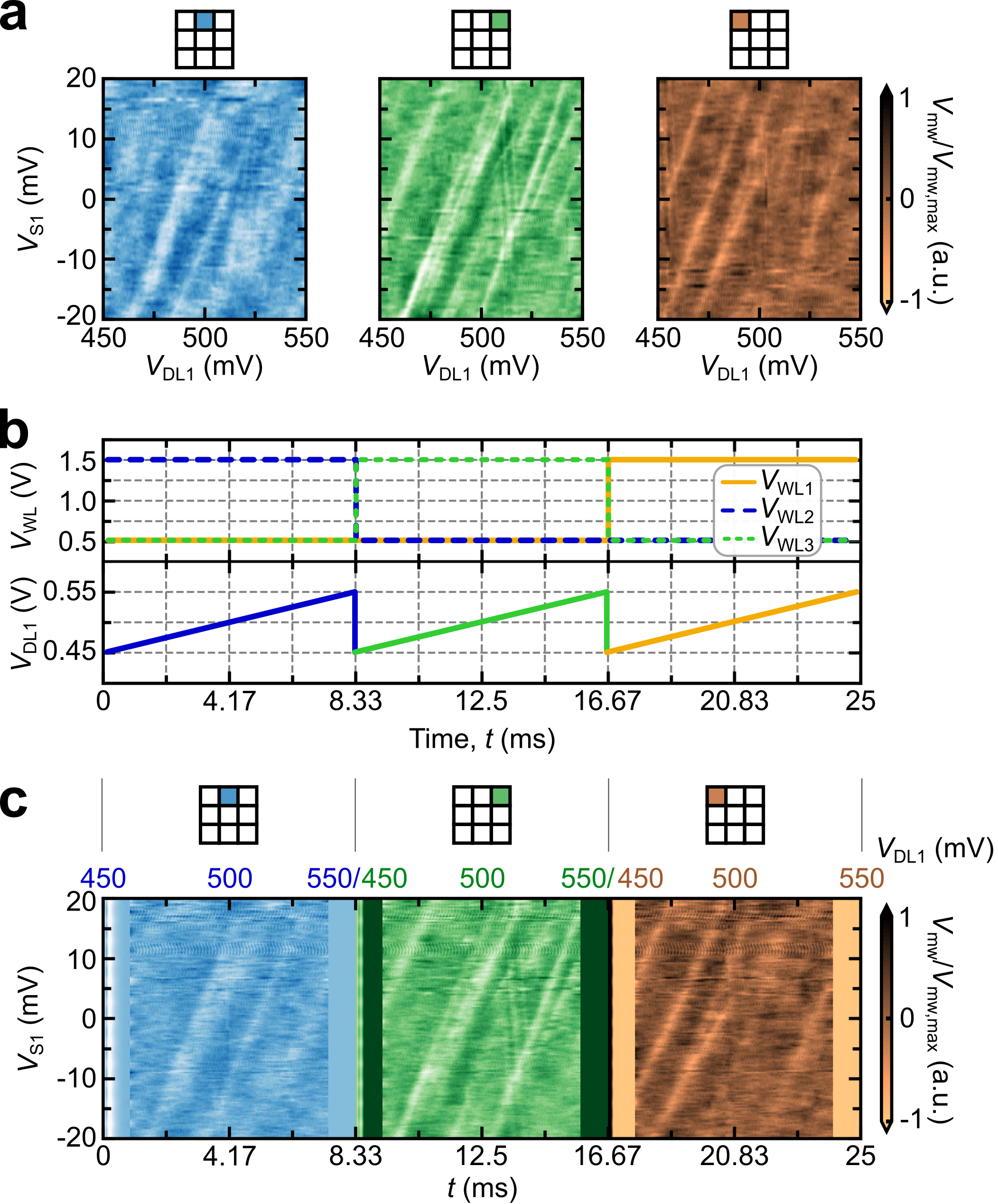}
\caption{\textbf{Integrated time-multiplexed readout.} \textbf{a,}  Individual gate-based reflectometry measurements for quantum devices $Q_\mathrm{12}$ (blue), $Q_\mathrm{13}$ (green), and $Q_\mathrm{11}$ (gold) with reflectometry signals $V_\mathrm{mw}$ as a function of $V_\mathrm{DL1}$ and $V_\mathrm{S1j}$. \textbf{b,} Sequences of $V_\mathrm{WLj}$ and $V_\mathrm{DLi}$ for time-domain multiplexing reflectometry sensing of $Q_\mathrm{12}$, $Q_\mathrm{13}$, $Q_\mathrm{11}$. $V_\mathrm{WL2}$ (blue), $V_\mathrm{WL3}$ (green), and $V_\mathrm{WL1}$ (gold) follow a square wave between 1.5~V and 0.5~V in time. $V_\mathrm{DL1}$ follows a ramping wave synchronized to $V_\mathrm{WLj}$ in time domain. \textbf{c,} Stability diagrams for $Q_\mathrm{12}$$\to$$Q_\mathrm{13}$$\to$$Q_\mathrm{11}$ in time domain corresponding to sequences of $V_\mathrm{WLj}$ and $V_\mathrm{DLi}$ in \textbf{b}. Measurements are performed at 50~mK and the carrier frequency is $f_\mathrm{mw1}$=6.872~GHz for the measurements in \textbf{a} and \textbf{c}. Data presented in \textbf{c} are processed by using the data processing method in Methods.}
\label{fig:tmux}
\end{figure}

\section*{Gate-based dispersive readout}

We use the resonators as sensors to perform integrated gate-based dispersive readout of the QD charge states. The resonators produce an oscillatory voltage on the gate of the QD which can result in cyclic tunneling of electrons back and forth the electronic reservoirs~\cite{ahmed2018commphys}. This results in an equivalent capacitance that modifies the impedance of the resonator producing a change in the reflected voltage ($V_\mathrm{mw}$).
\newline\indent
To benchmark the method, we performed DC transport measurements, shown in Fig.~\ref{fig:reflectometry}a, for device $Q_{13}$, and observe Coulomb peaks $\gamma$, $\delta$, and $\varepsilon$. We compare these results with those measured via reflectometry in the same $V_\mathrm{DLi}$ region (Fig.~\ref{fig:reflectometry}b), while exploring the dependence with the state of the access transistor. At low $V_\mathrm{WL3}-V_\mathrm{DL1}$ $<(0.786-0.340)$~V, the access transistor is highly resistive and the microwave signal is highly attenuated (\textit{Off} region). At intermediate $V_\mathrm{WL3}-V_\mathrm{DL1}$, the access transistor is in the depletion region and the oscillatory voltage at the transistor input produces changes in the capacitance that are picked-up as large changes in the reflected signal (\textit{Forbidden} region). Finally, at high $V_\mathrm{WL3}-V_\mathrm{DL1}$ $>(1.278-0.340)$~V, the access transistor presents a low resistance state and the microwave signal can travel through (\textit{On} region), exciting cyclic tunneling in the QD, which manifests as regions of enhanced $V_\mathrm{mw}$. Besides the same transitions as in Fig.~\ref{fig:reflectometry}a, two additional peaks, $\alpha$ and $\beta$, are observed. These are results of cyclic tunneling to one of the electron reservoirs only (as opposed to current, that requires sizable tunneling rates to both source and drain). This highlights the efficiency of gate-based sensing in detecting electronic transitions even if the QDs are offset from the center of the channel and present low tunnel rates to one ohmic contact. To provide further evidence of the nature of these transitions, we present individual $V_\mathrm{Sij}$-$V_\mathrm{DLi}$ maps in Fig.~\ref{fig:reflectometry}c. The dependence on $V_\mathrm{S}$ suggests that $\alpha$ and $\beta$ peaks are originated from cyclic tunneling to the source. A lineshape analysis (see Methods) from the data in Fig.~\ref{fig:reflectometry}c, reveals a signal-to-noise ratio (SNR) of 28.7 in 400~ms of integration time and a tunnel rate to the source of 48.3~GHz.
These measurements represent the first demonstration of fully-integrated conditional gate-based readout of quantum dots implemented in an industrial CMOS technology.

\section*{Time-multiplexed readout}

We then performed time-multiplexed reflectometry measurements of QD devices in the same row, by addressing them at the frequency of the shared resonator and activating the corresponding columns, one after the other. We chose $Resonator_1$ with a carrier frequency $f_\mathrm{mw1}$ = 6.872~GHz and used the 1$\times$3 quantum dot array [$Q_\mathrm{1j}$].
\newline\indent
We first measured the charge stability diagrams for each individual $Q_\mathrm{1j}$ (Fig.~\ref{fig:tmux}a), when the corresponding access transistor is active. We used such data as the control set, we then performed the dynamic characterization in time-domain multiplexing and compared the results. The dynamic voltage sequence consists of a digital high $V_\mathrm{WLj}$ voltage ($V_\mathrm{WLj}^{High}$=1.5~V) applied to the cell to be read, while the other two cells are set at low $V_\mathrm{WLj}$ voltages ($V_\mathrm{WLj}^{Low}$=0.5~V). The digital values are selected according to the \textit{On}-\textit{Off} regions in Fig.~\ref{fig:reflectometry}b. During that period, we simultaneously applied a voltage ramp to the data-line $V_\mathrm{DL1}$ to acquire the data from the $Q_\mathrm{1j}$ of the corresponding cell. We then sequentially raised the $V_\mathrm{WLj}$ voltage of the next cell while keeping the other two low. The full sequence is illustrated in Fig.~\ref{fig:tmux}b, where we first measured $Q_{12}$ followed by $Q_{13}$ and $Q_{11}$. Finally, we repeated the full sequence as we stepped the $V_\mathrm{S1j}$ voltages to acquire the charge stability maps in Fig.~\ref{fig:tmux}c. The match between the control set and the dynamic measurements indicates the success of the protocol.
\newline\indent
We note that time-domain multiplexing does not necessarily require sequential addressing but can be performed in a random-access manner similar to DRAM architectures. These results represent fully-integrated time-multiplexed reflectometry measurements of silicon QDs and demonstrate an important reduction in the analog readout infrastructure of quantum circuits, since multiple devices can be read out by a single resonator.

\begin{figure}[t]
\centering
\includegraphics[width=\columnwidth]{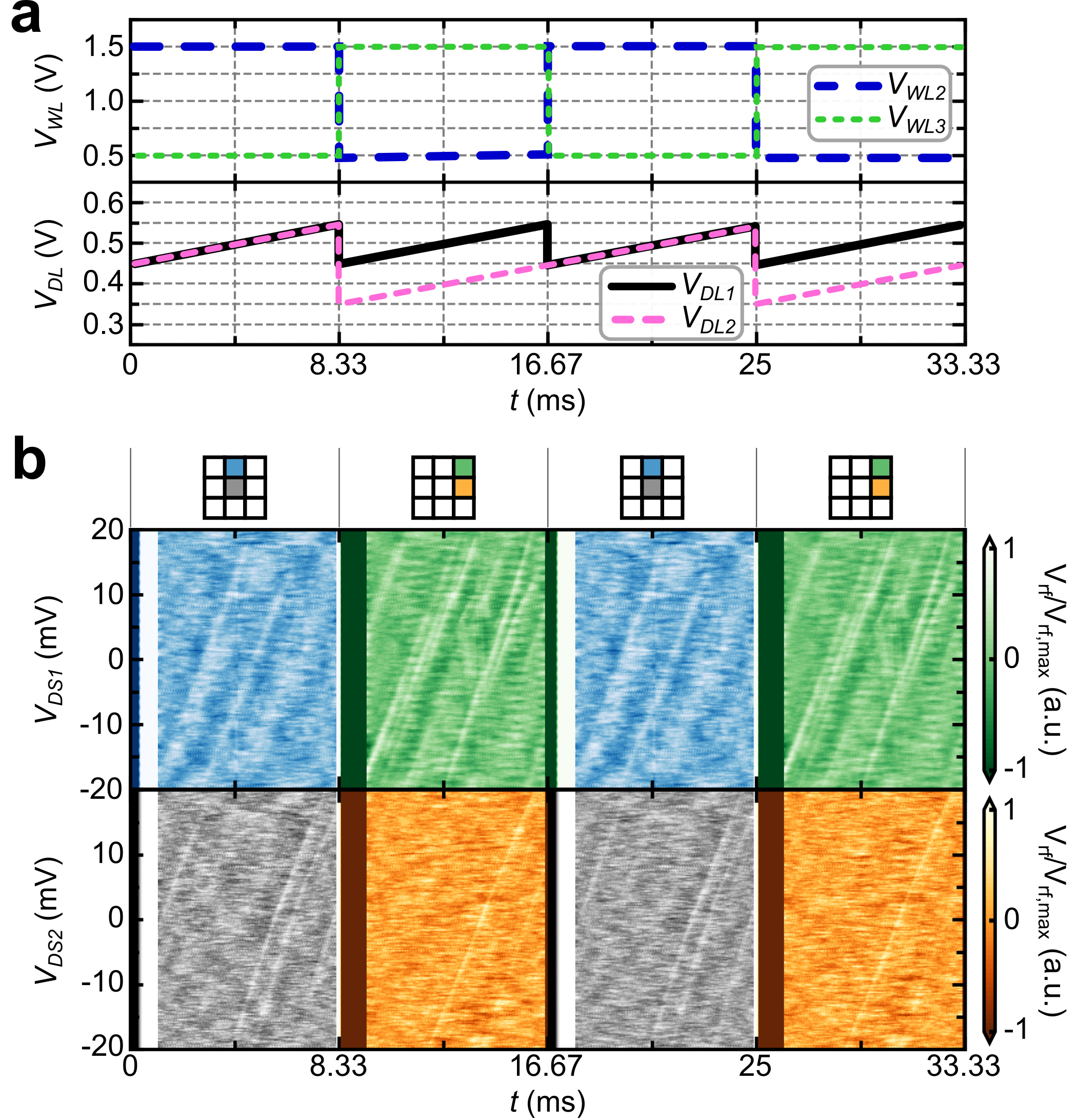}
\caption{\textbf{Integrated time- and frequency-multiplexed readout.} \textbf{a,} Sequence of $V_\mathrm{WLj}$ and $V_\mathrm{DLi}$ to perform time- and frequency-multiplexing reflectometry sensing using the 2$\times$2 sub-matrix [[$Q_\mathrm{12}$, $Q_\mathrm{13}$], [$Q_\mathrm{22}$, $Q_\mathrm{23}$]]. $V_\mathrm{WL2}$ (blue) and $V_\mathrm{WL3}$ (green) follow a square wave between 1.5~V and 0.5~V, while $V_\mathrm{WL1}$=0~V. $V_\mathrm{DL1}$ and $V_\mathrm{DL2}$ follow a ramping wave synchronized to $V_\mathrm{WL2}$ and $V_\mathrm{WL3}$ in time domain. \textbf{b,} Stability diagrams for quantum dot devices $Q_\mathrm{12}$ (blue)$\to$$Q_\mathrm{13}$ (green)$\to$$Q_\mathrm{12}$$\to$$Q_\mathrm{13}$ addressed through $Resonator_\mathrm{1}$ in time sequence, and for $Q_\mathrm{22}$ (grey)$\to$$Q_\mathrm{23}$ (orange)$\to$$Q_\mathrm{22}$$\to$$Q_\mathrm{23}$ simultaneously addressed through $Resonator_\mathrm{2}$. Measurements are performed at 50~mK and the carrier frequencies are $f_\mathrm{mw1}$=6.873~GHz and $f_\mathrm{mw2}$=7.419~GHz for $Resonator_\mathrm{1}$ and $Resonator_\mathrm{2}$, respectively.}
\label{fig:tfmux}
\end{figure}

\section*{Frequency-multiplexed readout}

Once time-multiplexing (addressing columns) in reflectometry has been demonstrated, using the second degree of freedom in the matrix, we demonstrate frequency-multiplexing (addressing rows). For this experiment, we used $Resonator_\mathrm{2}$ at $f_\mathrm{mw2}$=7.419~GHz with $Resonator_\mathrm{1}$ at $f_\mathrm{mw1}$=6.873~GHz, to perform parallel readout of two independent QDs on different rows (addressable at different frequencies). As shown in Fig.~\ref{fig:tfmux}a (left panels, up to 8.33~ms), the word-line voltage $V_\mathrm{WL2}$ was kept high to activate column 2, while both $V_\mathrm{DL1}$ and $V_\mathrm{DL2}$ were ramped up to activate row 1 and 2. As a result, a 2$\times$1 parallel set of Coulomb peak transitions from $Q_\mathrm{12}$ and $Q_\mathrm{22}$ was obtained (Fig.~\ref{fig:tfmux}b, left panels, up to 8.33~ms).
\newline\indent
Parallel readout reduces the overall integration time to read large quantum circuits by clustering in the same time step the readout of a number of devices. This feature is of particular relevance to quantum error correction sequences, such as the surface code, that requires continuous qubit readout at scale~\cite{fowler2012pra}.
These results demonstrate fully-integrated frequency-multiplexing gate-based readout of silicon quantum dots on a single chip and in the 6-8 GHz range.

\section*{Time- and frequency-multiplexed readout}

Finally, we combined time- and frequency domain multiplexing. $Resonator_\mathrm{1}$ and $Resonator_\mathrm{2}$ were used simultaneously for sensing the sub-arrays [$Q_\mathrm{12}$, $Q_\mathrm{13}$] and [$Q_\mathrm{22}$, $Q_\mathrm{23}$], respectively. Similar to the time-domain multiplexing readout, a sequence of square waves $V_\mathrm{WL2}$$\to$$V_\mathrm{WL3}$$\to$$V_\mathrm{WL2}$$\to$$V_\mathrm{WL3}$ was applied to the word-lines, as shown in Fig.~\ref{fig:tfmux}a, to selectively read out the transistors in different columns. Simultaneous arbitrary waves in $V_\mathrm{DL1}$ and $V_\mathrm{DL2}$ synchronized to $V_\mathrm{WL2}$ and $V_\mathrm{WL3}$, as shown in Fig.~\ref{fig:tfmux}a, were applied to read out two parallel rows. The result is a 2$\times$2 matrix of Coulomb maps from the addressed quantum dots (Fig.~\ref{fig:tfmux}b). The results presented here demonstrate, for the first time, the combination of time- and frequency-multiplexing gate-based readout for silicon QDs, in a fully-integrated platform, with a scalable architecture.

\section*{Conclusion}

We have presented a cryogenic IC fabricated using industrial 40-nm CMOS technology that contains three elements required in a silicon-based quantum computer: QDs, I/O management and multiplexed readout electronics. Our results probe the limits of integration using commercial CMOS technology and pave the way to the realization of an integrated silicon quantum processor. On the way forward, spin control will require attention. Single qubit control could be achieved by embedding the IC in a 3D cavity to perform electron spin resonance conditionally controlled by Stark shifts~\cite{laucht2015sa}. Two-qubit gates will require bringing the cells in close proximity to enable spin exchange interaction. Current industrial design rules for this process set the gate pitch limit to 120~nm, therefore some process customization will be required to bring it down to $\sim$70~nm, where sizeable tunnel coupling occurs~\cite{ansaloni2020natcomm}. Finally, the SNR could be improved by lowering the tunnel rates closer to the probing frequency, by increasing the quality factor of the inductors using industry-compatible superconductors (TiN)~\cite{shearrow2018apl} and by using traveling-wave or Josephson parametric amplification~\cite{schaal2020prl}. These improvements could reduce the minimum integration time by a factor of 100 or more, to bring it below the coherence time of the qubits~\cite{veldhorst2014natnano}, a necessity for error correction protocols.

\bibliography{references/references}

\begin{thebibliography}{47}%
\makeatletter
\providecommand \@ifxundefined [1]{%
 \@ifx{#1\undefined}
}%
\providecommand \@ifnum [1]{%
 \ifnum #1\expandafter \@firstoftwo
 \else \expandafter \@secondoftwo
 \fi
}%
\providecommand \@ifx [1]{%
 \ifx #1\expandafter \@firstoftwo
 \else \expandafter \@secondoftwo
 \fi
}%
\providecommand \natexlab [1]{#1}%
\providecommand \enquote  [1]{``#1''}%
\providecommand \bibnamefont  [1]{#1}%
\providecommand \bibfnamefont [1]{#1}%
\providecommand \citenamefont [1]{#1}%
\providecommand \href@noop [0]{\@secondoftwo}%
\providecommand \href [0]{\begingroup \@sanitize@url \@href}%
\providecommand \@href[1]{\@@startlink{#1}\@@href}%
\providecommand \@@href[1]{\endgroup#1\@@endlink}%
\providecommand \@sanitize@url [0]{\catcode `\\12\catcode `\$12\catcode
  `\&12\catcode `\#12\catcode `\^12\catcode `\_12\catcode `\%12\relax}%
\providecommand \@@startlink[1]{}%
\providecommand \@@endlink[0]{}%
\providecommand \url  [0]{\begingroup\@sanitize@url \@url }%
\providecommand \@url [1]{\endgroup\@href {#1}{\urlprefix }}%
\providecommand \urlprefix  [0]{URL }%
\providecommand \Eprint [0]{\href }%
\providecommand \doibase [0]{http://dx.doi.org/}%
\providecommand \selectlanguage [0]{\@gobble}%
\providecommand \bibinfo  [0]{\@secondoftwo}%
\providecommand \bibfield  [0]{\@secondoftwo}%
\providecommand \translation [1]{[#1]}%
\providecommand \BibitemOpen [0]{}%
\providecommand \bibitemStop [0]{}%
\providecommand \bibitemNoStop [0]{.\EOS\space}%
\providecommand \EOS [0]{\spacefactor3000\relax}%
\providecommand \BibitemShut  [1]{\csname bibitem#1\endcsname}%
\let\auto@bib@innerbib\@empty
%</preamble>
\bibitem [{\citenamefont {Arute}\ \emph {et~al.}(2019)\citenamefont {Arute}
  \emph {et~al.}}]{arute2019nature}%
  \BibitemOpen
  \bibfield  {author} {\bibinfo {author} {\bibfnamefont {F.}~\bibnamefont
  {Arute}} \emph {et~al.},\ }\bibfield  {title} {\enquote {\bibinfo {title}
  {Quantum supremacy using a programmable superconducting processor},}\ }\href
  {\doibase 10.1038/s41586-019-1666-5} {\bibfield  {journal} {\bibinfo
  {journal} {Nature}\ }\textbf {\bibinfo {volume} {574}},\ \bibinfo {pages}
  {505--510} (\bibinfo {year} {2019})}\BibitemShut {NoStop}%
\bibitem [{\citenamefont {Watson}\ \emph {et~al.}(2018)\citenamefont {Watson}
  \emph {et~al.}}]{watson2018nature}%
  \BibitemOpen
  \bibfield  {author} {\bibinfo {author} {\bibfnamefont {T.~F.}\ \bibnamefont
  {Watson}} \emph {et~al.},\ }\bibfield  {title} {\enquote {\bibinfo {title} {A
  programmable two-qubit quantum processor in silicon},}\ }\href {\doibase
  10.1038/nature25766} {\bibfield  {journal} {\bibinfo  {journal} {Nature}\
  }\textbf {\bibinfo {volume} {555}},\ \bibinfo {pages} {633--637} (\bibinfo
  {year} {2018})}\BibitemShut {NoStop}%
\bibitem [{\citenamefont {Reilly}(2015)}]{reilly2015npjqi}%
  \BibitemOpen
  \bibfield  {author} {\bibinfo {author} {\bibfnamefont {D.~J.}\ \bibnamefont
  {Reilly}},\ }\bibfield  {title} {\enquote {\bibinfo {title} {Engineering the
  quantum-classical interface of solid-state qubits},}\ }\href {\doibase
  10.1038/npjqi.2015.11} {\bibfield  {journal} {\bibinfo  {journal} {npj
  Quantum Information}\ }\textbf {\bibinfo {volume} {1}},\ \bibinfo {pages}
  {15011} (\bibinfo {year} {2015})}\BibitemShut {NoStop}%
\bibitem [{\citenamefont {Charbon}\ \emph {et~al.}(2016)\citenamefont {Charbon}
  \emph {et~al.}}]{charbon2016iedm}%
  \BibitemOpen
  \bibfield  {author} {\bibinfo {author} {\bibfnamefont {E.}~\bibnamefont
  {Charbon}} \emph {et~al.},\ }\bibfield  {title} {\enquote {\bibinfo {title}
  {Cryo-{CMOS} for quantum computing},}\ }in\ \href {\doibase
  10.1109/IEDM.2016.7838410} {\emph {\bibinfo {booktitle} {2016 IEEE
  International Electron Devices Meeting (IEDM)}}}\ (\bibinfo {year} {2016})\
  pp.\ \bibinfo {pages} {13.5.1--13.5.4}\BibitemShut {NoStop}%
\bibitem [{\citenamefont {{Reilly}}(2019)}]{reilly2019iedm}%
  \BibitemOpen
  \bibfield  {author} {\bibinfo {author} {\bibfnamefont {D.~J.}\ \bibnamefont
  {{Reilly}}},\ }\bibfield  {title} {\enquote {\bibinfo {title} {{Challenges in
  Scaling-up the Control Interface of a Quantum Computer}},}\ }in\ \href
  {\doibase 10.1109/IEDM19573.2019.8993497} {\emph {\bibinfo {booktitle} {2019
  IEEE International Electron Devices Meeting (IEDM)}}}\ (\bibinfo {year}
  {2019})\ pp.\ \bibinfo {pages} {31.7.1--31.7.6}\BibitemShut {NoStop}%
\bibitem [{\citenamefont {Maurand}\ \emph {et~al.}(2016)\citenamefont {Maurand}
  \emph {et~al.}}]{maurand2016natcomm}%
  \BibitemOpen
  \bibfield  {author} {\bibinfo {author} {\bibfnamefont {R.}~\bibnamefont
  {Maurand}} \emph {et~al.},\ }\bibfield  {title} {\enquote {\bibinfo {title}
  {{A CMOS silicon spin qubit}},}\ }\href {\doibase 10.1038/ncomms13575}
  {\bibfield  {journal} {\bibinfo  {journal} {Nature Communications}\ }\textbf
  {\bibinfo {volume} {7}},\ \bibinfo {pages} {13575} (\bibinfo {year}
  {2016})}\BibitemShut {NoStop}%
\bibitem [{\citenamefont {Montanaro}(2016)}]{montanaro2016npjqi}%
  \BibitemOpen
  \bibfield  {author} {\bibinfo {author} {\bibfnamefont {A.}~\bibnamefont
  {Montanaro}},\ }\bibfield  {title} {\enquote {\bibinfo {title} {Quantum
  algorithms: an overview},}\ }\href {\doibase 10.1038/npjqi.2015.23}
  {\bibfield  {journal} {\bibinfo  {journal} {npj Quantum Information}\
  }\textbf {\bibinfo {volume} {2}},\ \bibinfo {pages} {15023} (\bibinfo {year}
  {2016})}\BibitemShut {NoStop}%
\bibitem [{\citenamefont {Veldhorst}\ \emph {et~al.}(2014)\citenamefont
  {Veldhorst} \emph {et~al.}}]{veldhorst2014natnano}%
  \BibitemOpen
  \bibfield  {author} {\bibinfo {author} {\bibfnamefont {M.}~\bibnamefont
  {Veldhorst}} \emph {et~al.},\ }\bibfield  {title} {\enquote {\bibinfo {title}
  {An addressable quantum dot qubit with fault-tolerant control-fidelity},}\
  }\href {\doibase 10.1038/nnano.2014.216} {\bibfield  {journal} {\bibinfo
  {journal} {Nature Nanotechnology}\ }\textbf {\bibinfo {volume} {9}},\
  \bibinfo {pages} {981--985} (\bibinfo {year} {2014})}\BibitemShut {NoStop}%
\bibitem [{\citenamefont {Urdampilleta}\ \emph {et~al.}(2019)\citenamefont
  {Urdampilleta} \emph {et~al.}}]{urdampilleta2019natnano}%
  \BibitemOpen
  \bibfield  {author} {\bibinfo {author} {\bibfnamefont {M.}~\bibnamefont
  {Urdampilleta}} \emph {et~al.},\ }\bibfield  {title} {\enquote {\bibinfo
  {title} {Gate-based high fidelity spin readout in a {CMOS} device},}\ }\href
  {\doibase 10.1038/s41565-019-0443-9} {\bibfield  {journal} {\bibinfo
  {journal} {Nature Nanotechnology}\ }\textbf {\bibinfo {volume} {14}},\
  \bibinfo {pages} {737--741} (\bibinfo {year} {2019})}\BibitemShut {NoStop}%
\bibitem [{\citenamefont {Yoneda}\ \emph {et~al.}(2018)\citenamefont {Yoneda}
  \emph {et~al.}}]{yoneda2018natnano}%
  \BibitemOpen
  \bibfield  {author} {\bibinfo {author} {\bibfnamefont {J.}~\bibnamefont
  {Yoneda}} \emph {et~al.},\ }\bibfield  {title} {\enquote {\bibinfo {title} {A
  quantum-dot spin qubit with coherence limited by charge noise and fidelity
  higher than 99.9{\%}},}\ }\href {\doibase 10.1038/s41565-017-0014-x}
  {\bibfield  {journal} {\bibinfo  {journal} {Nature Nanotechnology}\ }\textbf
  {\bibinfo {volume} {13}},\ \bibinfo {pages} {102--106} (\bibinfo {year}
  {2018})}\BibitemShut {NoStop}%
\bibitem [{\citenamefont {Huang}\ \emph {et~al.}(2019)\citenamefont {Huang}
  \emph {et~al.}}]{huang2019nature}%
  \BibitemOpen
  \bibfield  {author} {\bibinfo {author} {\bibfnamefont {W.}~\bibnamefont
  {Huang}} \emph {et~al.},\ }\bibfield  {title} {\enquote {\bibinfo {title}
  {Fidelity benchmarks for two-qubit gates in silicon},}\ }\href {\doibase
  10.1038/s41586-019-1197-0} {\bibfield  {journal} {\bibinfo  {journal}
  {Nature}\ }\textbf {\bibinfo {volume} {569}},\ \bibinfo {pages} {532--536}
  (\bibinfo {year} {2019})}\BibitemShut {NoStop}%
\bibitem [{\citenamefont {Zajac}\ \emph {et~al.}(2018)\citenamefont {Zajac}
  \emph {et~al.}}]{zajac2018science}%
  \BibitemOpen
  \bibfield  {author} {\bibinfo {author} {\bibfnamefont {D.~M.}\ \bibnamefont
  {Zajac}} \emph {et~al.},\ }\bibfield  {title} {\enquote {\bibinfo {title}
  {Resonantly driven {CNOT} gate for electron spins},}\ }\href {\doibase
  10.1126/science.aao5965} {\bibfield  {journal} {\bibinfo  {journal}
  {Science}\ }\textbf {\bibinfo {volume} {359}},\ \bibinfo {pages} {439--442}
  (\bibinfo {year} {2018})}\BibitemShut {NoStop}%
\bibitem [{\citenamefont {Fowler}\ \emph {et~al.}(2012)\citenamefont {Fowler}
  \emph {et~al.}}]{fowler2012pra}%
  \BibitemOpen
  \bibfield  {author} {\bibinfo {author} {\bibfnamefont {A.~G.}\ \bibnamefont
  {Fowler}} \emph {et~al.},\ }\bibfield  {title} {\enquote {\bibinfo {title}
  {Surface codes: {Towards} practical large-scale quantum computation},}\
  }\href {\doibase 10.1103/PhysRevA.86.032324} {\bibfield  {journal} {\bibinfo
  {journal} {Physical Review A}\ }\textbf {\bibinfo {volume} {86}},\ \bibinfo
  {pages} {032324} (\bibinfo {year} {2012})}\BibitemShut {NoStop}%
\bibitem [{\citenamefont {Kawakami}\ \emph {et~al.}(2014)\citenamefont
  {Kawakami} \emph {et~al.}}]{kawakami2014natnano}%
  \BibitemOpen
  \bibfield  {author} {\bibinfo {author} {\bibfnamefont {E.}~\bibnamefont
  {Kawakami}} \emph {et~al.},\ }\bibfield  {title} {\enquote {\bibinfo {title}
  {Electrical control of a long-lived spin qubit in a {Si}/{SiGe} quantum
  dot},}\ }\href {\doibase 10.1038/nnano.2014.153} {\bibfield  {journal}
  {\bibinfo  {journal} {Nature Nanotechnology}\ }\textbf {\bibinfo {volume}
  {9}},\ \bibinfo {pages} {666--670} (\bibinfo {year} {2014})}\BibitemShut
  {NoStop}%
\bibitem [{\citenamefont {Yang}\ \emph
  {et~al.}(2020{\natexlab{a}})\citenamefont {Yang} \emph
  {et~al.}}]{yang2020edl}%
  \BibitemOpen
  \bibfield  {author} {\bibinfo {author} {\bibfnamefont {T.-Y.}\ \bibnamefont
  {Yang}} \emph {et~al.},\ }\bibfield  {title} {\enquote {\bibinfo {title}
  {{Quantum Transport in 40-nm MOSFETs at Deep-Cryogenic Temperatures}},}\
  }\href {\doibase 10.1109/LED.2020.2995645} {\bibfield  {journal} {\bibinfo
  {journal} {IEEE Electron Device Letters}\ }\textbf {\bibinfo {volume} {41}},\
  \bibinfo {pages} {981--984} (\bibinfo {year}
  {2020}{\natexlab{a}})}\BibitemShut {NoStop}%
\bibitem [{\citenamefont {{Bonen}}\ \emph {et~al.}(2019)\citenamefont {{Bonen}}
  \emph {et~al.}}]{bonen2019edl}%
  \BibitemOpen
  \bibfield  {author} {\bibinfo {author} {\bibfnamefont {S.}~\bibnamefont
  {{Bonen}}} \emph {et~al.},\ }\bibfield  {title} {\enquote {\bibinfo {title}
  {{Cryogenic Characterization of 22-nm FDSOI CMOS Technology for Quantum
  Computing ICs}},}\ }\href {\doibase 10.1109/LED.2018.2880303} {\bibfield
  {journal} {\bibinfo  {journal} {IEEE Electron Device Letters}\ }\textbf
  {\bibinfo {volume} {40}},\ \bibinfo {pages} {127--130} (\bibinfo {year}
  {2019})}\BibitemShut {NoStop}%
\bibitem [{\citenamefont {Veldhorst}\ \emph {et~al.}(2017)\citenamefont
  {Veldhorst} \emph {et~al.}}]{veldhorst2017natcomm}%
  \BibitemOpen
  \bibfield  {author} {\bibinfo {author} {\bibfnamefont {M.}~\bibnamefont
  {Veldhorst}} \emph {et~al.},\ }\bibfield  {title} {\enquote {\bibinfo {title}
  {{Silicon CMOS architecture for a spin-based quantum computer}},}\ }\href
  {\doibase 10.1038/s41467-017-01905-6} {\bibfield  {journal} {\bibinfo
  {journal} {Nature Communications}\ }\textbf {\bibinfo {volume} {8}},\
  \bibinfo {pages} {1766} (\bibinfo {year} {2017})}\BibitemShut {NoStop}%
\bibitem [{\citenamefont {Vandersypen}\ \emph {et~al.}(2017)\citenamefont
  {Vandersypen} \emph {et~al.}}]{vandersypen2017npjqi}%
  \BibitemOpen
  \bibfield  {author} {\bibinfo {author} {\bibfnamefont {L.~M.~K.}\
  \bibnamefont {Vandersypen}} \emph {et~al.},\ }\bibfield  {title} {\enquote
  {\bibinfo {title} {Interfacing spin qubits in quantum dots and donors-hot,
  dense, and coherent},}\ }\href {\doibase 10.1038/s41534-017-0038-y}
  {\bibfield  {journal} {\bibinfo  {journal} {npj Quantum Information}\
  }\textbf {\bibinfo {volume} {3}},\ \bibinfo {pages} {34} (\bibinfo {year}
  {2017})}\BibitemShut {NoStop}%
\bibitem [{\citenamefont {Li}\ \emph {et~al.}(2018)\citenamefont {Li} \emph
  {et~al.}}]{li2018sa}%
  \BibitemOpen
  \bibfield  {author} {\bibinfo {author} {\bibfnamefont {R.}~\bibnamefont {Li}}
  \emph {et~al.},\ }\bibfield  {title} {\enquote {\bibinfo {title} {A crossbar
  network for silicon quantum dot qubits},}\ }\href {\doibase
  10.1126/sciadv.aar3960} {\bibfield  {journal} {\bibinfo  {journal} {Science
  Advances}\ }\textbf {\bibinfo {volume} {4}},\ \bibinfo {pages} {3960}
  (\bibinfo {year} {2018})}\BibitemShut {NoStop}%
\bibitem [{\citenamefont {{Le Guevel}}\ \emph {et~al.}(2020)\citenamefont {{Le
  Guevel}} \emph {et~al.}}]{leguevel2020isscc}%
  \BibitemOpen
  \bibfield  {author} {\bibinfo {author} {\bibfnamefont {L.}~\bibnamefont {{Le
  Guevel}}} \emph {et~al.},\ }\bibfield  {title} {\enquote {\bibinfo {title}
  {{A 110mK 295$\upmu$W 28nm FDSOI CMOS Quantum Integrated Circuit with a
  2.8GHz Excitation and nA Current Sensing of an On-Chip Double Quantum
  Dot}},}\ }in\ \href {\doibase 10.1109/ISSCC19947.2020.9063090} {\emph
  {\bibinfo {booktitle} {2020 IEEE International Solid-State Circuits
  Conference - (ISSCC)}}}\ (\bibinfo {year} {2020})\ pp.\ \bibinfo {pages}
  {306--308}\BibitemShut {NoStop}%
\bibitem [{\citenamefont {Opremcak}\ \emph {et~al.}(2018)\citenamefont
  {Opremcak} \emph {et~al.}}]{opremcak2018science}%
  \BibitemOpen
  \bibfield  {author} {\bibinfo {author} {\bibfnamefont {A.}~\bibnamefont
  {Opremcak}} \emph {et~al.},\ }\bibfield  {title} {\enquote {\bibinfo {title}
  {Measurement of a superconducting qubit with a microwave photon counter},}\
  }\href {\doibase 10.1126/science.aat4625} {\bibfield  {journal} {\bibinfo
  {journal} {Science}\ }\textbf {\bibinfo {volume} {361}},\ \bibinfo {pages}
  {1239--1242} (\bibinfo {year} {2018})}\BibitemShut {NoStop}%
\bibitem [{\citenamefont {Yang}\ \emph
  {et~al.}(2020{\natexlab{b}})\citenamefont {Yang} \emph
  {et~al.}}]{yang2020nature}%
  \BibitemOpen
  \bibfield  {author} {\bibinfo {author} {\bibfnamefont {C.~H.}\ \bibnamefont
  {Yang}} \emph {et~al.},\ }\bibfield  {title} {\enquote {\bibinfo {title}
  {{Operation of a silicon quantum processor unit cell above one Kelvin}},}\
  }\href {\doibase 10.1038/s41586-020-2171-6} {\bibfield  {journal} {\bibinfo
  {journal} {Nature}\ }\textbf {\bibinfo {volume} {580}},\ \bibinfo {pages}
  {350--354} (\bibinfo {year} {2020}{\natexlab{b}})}\BibitemShut {NoStop}%
\bibitem [{\citenamefont {Petit}\ \emph {et~al.}(2020)\citenamefont {Petit}
  \emph {et~al.}}]{petit2020nature}%
  \BibitemOpen
  \bibfield  {author} {\bibinfo {author} {\bibfnamefont {L.}~\bibnamefont
  {Petit}} \emph {et~al.},\ }\bibfield  {title} {\enquote {\bibinfo {title}
  {Universal quantum logic in hot silicon qubits},}\ }\href {\doibase
  10.1038/s41586-020-2170-7} {\bibfield  {journal} {\bibinfo  {journal}
  {Nature}\ }\textbf {\bibinfo {volume} {580}},\ \bibinfo {pages} {355--359}
  (\bibinfo {year} {2020})}\BibitemShut {NoStop}%
\bibitem [{\citenamefont {Xue}\ \emph {et~al.}(2020)\citenamefont {Xue} \emph
  {et~al.}}]{xue2020arxiv}%
  \BibitemOpen
  \bibfield  {author} {\bibinfo {author} {\bibfnamefont {X.}~\bibnamefont
  {Xue}} \emph {et~al.},\ }\href@noop {} {\enquote {\bibinfo {title}
  {{CMOS-based cryogenic control of silicon quantum circuits}},}\ } (\bibinfo
  {year} {2020}),\ \Eprint {http://arxiv.org/abs/2009.14185} {arXiv:2009.14185
  [quant-ph]} \BibitemShut {NoStop}%
\bibitem [{\citenamefont {Pauka}\ \emph {et~al.}(2019)\citenamefont {Pauka}
  \emph {et~al.}}]{pauka2019arxiv}%
  \BibitemOpen
  \bibfield  {author} {\bibinfo {author} {\bibfnamefont {S.~J.}\ \bibnamefont
  {Pauka}} \emph {et~al.},\ }\href@noop {} {\enquote {\bibinfo {title} {{A
  Cryogenic Interface for Controlling Many Qubits}},}\ } (\bibinfo {year}
  {2019}),\ \Eprint {http://arxiv.org/abs/1912.01299} {arXiv:1912.01299
  [quant-ph]} \BibitemShut {NoStop}%
\bibitem [{\citenamefont {Hornibrook}\ \emph {et~al.}(2014)\citenamefont
  {Hornibrook} \emph {et~al.}}]{hornibrook2014apl}%
  \BibitemOpen
  \bibfield  {author} {\bibinfo {author} {\bibfnamefont {J.~M.}\ \bibnamefont
  {Hornibrook}} \emph {et~al.},\ }\bibfield  {title} {\enquote {\bibinfo
  {title} {Frequency multiplexing for readout of spin qubits},}\ }\href
  {\doibase 10.1063/1.4868107} {\bibfield  {journal} {\bibinfo  {journal}
  {Applied Physics Letters}\ }\textbf {\bibinfo {volume} {104}},\ \bibinfo
  {pages} {103108} (\bibinfo {year} {2014})}\BibitemShut {NoStop}%
\bibitem [{\citenamefont {Pauka}\ \emph {et~al.}(2020)\citenamefont {Pauka}
  \emph {et~al.}}]{pauka2020pra}%
  \BibitemOpen
  \bibfield  {author} {\bibinfo {author} {\bibfnamefont {S.~J.}\ \bibnamefont
  {Pauka}} \emph {et~al.},\ }\bibfield  {title} {\enquote {\bibinfo {title}
  {{Characterizing Quantum Devices at Scale with Custom Cryo-CMOS}},}\ }\href
  {\doibase 10.1103/PhysRevApplied.13.054072} {\bibfield  {journal} {\bibinfo
  {journal} {Phys. Rev. Applied}\ }\textbf {\bibinfo {volume} {13}},\ \bibinfo
  {pages} {054072} (\bibinfo {year} {2020})}\BibitemShut {NoStop}%
\bibitem [{\citenamefont {Poto{\v{c}}nik}\ \emph {et~al.}(2020)\citenamefont
  {Poto{\v{c}}nik} \emph {et~al.}}]{potocnik2020arxiv}%
  \BibitemOpen
  \bibfield  {author} {\bibinfo {author} {\bibfnamefont {A.}~\bibnamefont
  {Poto{\v{c}}nik}} \emph {et~al.},\ }\href@noop {} {\enquote {\bibinfo {title}
  {{Millikelvin temperature cryo-CMOS multiplexer for scalable quantum device
  characterisation}},}\ } (\bibinfo {year} {2020}),\ \Eprint
  {http://arxiv.org/abs/2011.11514} {arXiv:2011.11514 [quant-ph]} \BibitemShut
  {NoStop}%
\bibitem [{\citenamefont {Paquelet~Wuetz}\ \emph {et~al.}(2020)\citenamefont
  {Paquelet~Wuetz} \emph {et~al.}}]{paqueletwuetz2020npjqi}%
  \BibitemOpen
  \bibfield  {author} {\bibinfo {author} {\bibfnamefont {B.}~\bibnamefont
  {Paquelet~Wuetz}} \emph {et~al.},\ }\bibfield  {title} {\enquote {\bibinfo
  {title} {{Multiplexed quantum transport using commercial off-the-shelf CMOS
  at sub-Kelvin temperatures}},}\ }\href {\doibase 10.1038/s41534-020-0274-4}
  {\bibfield  {journal} {\bibinfo  {journal} {npj Quantum Information}\
  }\textbf {\bibinfo {volume} {6}},\ \bibinfo {pages} {43} (\bibinfo {year}
  {2020})}\BibitemShut {NoStop}%
\bibitem [{\citenamefont {Schaal}\ \emph {et~al.}(2019)\citenamefont {Schaal}
  \emph {et~al.}}]{schaal2019natel}%
  \BibitemOpen
  \bibfield  {author} {\bibinfo {author} {\bibfnamefont {S.}~\bibnamefont
  {Schaal}} \emph {et~al.},\ }\bibfield  {title} {\enquote {\bibinfo {title}
  {{A CMOS dynamic random access architecture for radio-frequency readout of
  quantum devices}},}\ }\href {\doibase 10.1038/s41928-019-0259-5} {\bibfield
  {journal} {\bibinfo  {journal} {Nature Electronics}\ }\textbf {\bibinfo
  {volume} {2}},\ \bibinfo {pages} {236--242} (\bibinfo {year}
  {2019})}\BibitemShut {NoStop}%
\bibitem [{\citenamefont {West}\ \emph {et~al.}(2019)\citenamefont {West} \emph
  {et~al.}}]{west2019natnano}%
  \BibitemOpen
  \bibfield  {author} {\bibinfo {author} {\bibfnamefont {A.}~\bibnamefont
  {West}} \emph {et~al.},\ }\bibfield  {title} {\enquote {\bibinfo {title}
  {Gate-based single-shot readout of spins in silicon},}\ }\href {\doibase
  10.1038/s41565-019-0400-7} {\bibfield  {journal} {\bibinfo  {journal} {Nature
  Nanotechnology}\ }\textbf {\bibinfo {volume} {14}},\ \bibinfo {pages}
  {437--441} (\bibinfo {year} {2019})}\BibitemShut {NoStop}%
\bibitem [{\citenamefont {Colless}\ \emph {et~al.}(2013)\citenamefont {Colless}
  \emph {et~al.}}]{colless2013prl}%
  \BibitemOpen
  \bibfield  {author} {\bibinfo {author} {\bibfnamefont {J.~I.}\ \bibnamefont
  {Colless}} \emph {et~al.},\ }\bibfield  {title} {\enquote {\bibinfo {title}
  {{Dispersive Readout of a Few-Electron Double Quantum Dot with Fast rf Gate
  Sensors}},}\ }\href {\doibase 10.1103/PhysRevLett.110.046805} {\bibfield
  {journal} {\bibinfo  {journal} {Phys. Rev. Lett.}\ }\textbf {\bibinfo
  {volume} {110}},\ \bibinfo {pages} {046805} (\bibinfo {year}
  {2013})}\BibitemShut {NoStop}%
\bibitem [{\citenamefont {Wallraff}\ \emph {et~al.}(2004)\citenamefont
  {Wallraff} \emph {et~al.}}]{wallraff2004nature}%
  \BibitemOpen
  \bibfield  {author} {\bibinfo {author} {\bibfnamefont {A.}~\bibnamefont
  {Wallraff}} \emph {et~al.},\ }\bibfield  {title} {\enquote {\bibinfo {title}
  {Strong coupling of a single photon to a superconducting qubit using circuit
  quantum electrodynamics},}\ }\href {\doibase 10.1038/nature02851} {\bibfield
  {journal} {\bibinfo  {journal} {Nature}\ }\textbf {\bibinfo {volume} {431}},\
  \bibinfo {pages} {162--167} (\bibinfo {year} {2004})}\BibitemShut {NoStop}%
\bibitem [{\citenamefont {Zwanenburg}\ \emph {et~al.}(2013)\citenamefont
  {Zwanenburg} \emph {et~al.}}]{zwanenburg2013rmp}%
  \BibitemOpen
  \bibfield  {author} {\bibinfo {author} {\bibfnamefont {F.~A.}\ \bibnamefont
  {Zwanenburg}} \emph {et~al.},\ }\bibfield  {title} {\enquote {\bibinfo
  {title} {Silicon quantum electronics},}\ }\href {\doibase
  10.1103/RevModPhys.85.961} {\bibfield  {journal} {\bibinfo  {journal} {Rev.
  Mod. Phys.}\ }\textbf {\bibinfo {volume} {85}},\ \bibinfo {pages} {961--1019}
  (\bibinfo {year} {2013})}\BibitemShut {NoStop}%
\bibitem [{\citenamefont {Ares}\ \emph {et~al.}(2016)\citenamefont {Ares} \emph
  {et~al.}}]{ares2016pra}%
  \BibitemOpen
  \bibfield  {author} {\bibinfo {author} {\bibfnamefont {N.}~\bibnamefont
  {Ares}} \emph {et~al.},\ }\bibfield  {title} {\enquote {\bibinfo {title}
  {{Sensitive Radio-Frequency Measurements of a Quantum Dot by Tuning to
  Perfect Impedance Matching}},}\ }\href {\doibase
  10.1103/PhysRevApplied.5.034011} {\bibfield  {journal} {\bibinfo  {journal}
  {Phys. Rev. Applied}\ }\textbf {\bibinfo {volume} {5}},\ \bibinfo {pages}
  {034011} (\bibinfo {year} {2016})}\BibitemShut {NoStop}%
\bibitem [{\citenamefont {Samkharadze}\ \emph {et~al.}(2018)\citenamefont
  {Samkharadze} \emph {et~al.}}]{samkharadze2018science}%
  \BibitemOpen
  \bibfield  {author} {\bibinfo {author} {\bibfnamefont {N.}~\bibnamefont
  {Samkharadze}} \emph {et~al.},\ }\bibfield  {title} {\enquote {\bibinfo
  {title} {Strong spin-photon coupling in silicon},}\ }\href {\doibase
  10.1126/science.aar4054} {\bibfield  {journal} {\bibinfo  {journal}
  {Science}\ }\textbf {\bibinfo {volume} {359}},\ \bibinfo {pages} {1123--1127}
  (\bibinfo {year} {2018})}\BibitemShut {NoStop}%
\bibitem [{\citenamefont {Mi}\ \emph {et~al.}(2018)\citenamefont {Mi} \emph
  {et~al.}}]{mi2018nature}%
  \BibitemOpen
  \bibfield  {author} {\bibinfo {author} {\bibfnamefont {X.}~\bibnamefont {Mi}}
  \emph {et~al.},\ }\bibfield  {title} {\enquote {\bibinfo {title} {A coherent
  spin-photon interface in silicon},}\ }\href {\doibase 10.1038/nature25769}
  {\bibfield  {journal} {\bibinfo  {journal} {Nature}\ }\textbf {\bibinfo
  {volume} {555}},\ \bibinfo {pages} {599--603} (\bibinfo {year}
  {2018})}\BibitemShut {NoStop}%
\bibitem [{\citenamefont {Zheng}\ \emph {et~al.}(2019)\citenamefont {Zheng}
  \emph {et~al.}}]{zheng2019natnano}%
  \BibitemOpen
  \bibfield  {author} {\bibinfo {author} {\bibfnamefont {G.}~\bibnamefont
  {Zheng}} \emph {et~al.},\ }\bibfield  {title} {\enquote {\bibinfo {title}
  {Rapid gate-based spin read-out in silicon using an on-chip resonator},}\
  }\href {\doibase 10.1038/s41565-019-0488-9} {\bibfield  {journal} {\bibinfo
  {journal} {Nature Nanotechnology}\ }\textbf {\bibinfo {volume} {14}},\
  \bibinfo {pages} {742--746} (\bibinfo {year} {2019})}\BibitemShut {NoStop}%
\bibitem [{\citenamefont {Ahmed}\ \emph
  {et~al.}(2018{\natexlab{a}})\citenamefont {Ahmed} \emph
  {et~al.}}]{ahmed2018pra}%
  \BibitemOpen
  \bibfield  {author} {\bibinfo {author} {\bibfnamefont {I.}~\bibnamefont
  {Ahmed}} \emph {et~al.},\ }\bibfield  {title} {\enquote {\bibinfo {title}
  {{Radio-Frequency Capacitive Gate-Based Sensing}},}\ }\href {\doibase
  10.1103/PhysRevApplied.10.014018} {\bibfield  {journal} {\bibinfo  {journal}
  {Phys. Rev. Applied}\ }\textbf {\bibinfo {volume} {10}},\ \bibinfo {pages}
  {014018} (\bibinfo {year} {2018}{\natexlab{a}})}\BibitemShut {NoStop}%
\bibitem [{\citenamefont {Gonzalez-Zalba}\ \emph {et~al.}(2015)\citenamefont
  {Gonzalez-Zalba} \emph {et~al.}}]{gonzalez-zalba2015natcomm}%
  \BibitemOpen
  \bibfield  {author} {\bibinfo {author} {\bibfnamefont {M.~F.}\ \bibnamefont
  {Gonzalez-Zalba}} \emph {et~al.},\ }\bibfield  {title} {\enquote {\bibinfo
  {title} {Probing the limits of gate-based charge sensing},}\ }\href {\doibase
  10.1038/ncomms7084} {\bibfield  {journal} {\bibinfo  {journal} {Nature
  Communications}\ }\textbf {\bibinfo {volume} {6}},\ \bibinfo {pages} {6084}
  (\bibinfo {year} {2015})}\BibitemShut {NoStop}%
\bibitem [{\citenamefont {Ahmed}\ \emph
  {et~al.}(2018{\natexlab{b}})\citenamefont {Ahmed} \emph
  {et~al.}}]{ahmed2018commphys}%
  \BibitemOpen
  \bibfield  {author} {\bibinfo {author} {\bibfnamefont {I.}~\bibnamefont
  {Ahmed}} \emph {et~al.},\ }\bibfield  {title} {\enquote {\bibinfo {title}
  {Primary thermometry of a single reservoir using cyclic electron tunneling to
  a quantum dot},}\ }\href {\doibase 10.1038/s42005-018-0066-8} {\bibfield
  {journal} {\bibinfo  {journal} {Communications Physics}\ }\textbf {\bibinfo
  {volume} {1}},\ \bibinfo {pages} {66} (\bibinfo {year}
  {2018}{\natexlab{b}})}\BibitemShut {NoStop}%
\bibitem [{\citenamefont {Laucht}\ \emph {et~al.}(2015)\citenamefont {Laucht}
  \emph {et~al.}}]{laucht2015sa}%
  \BibitemOpen
  \bibfield  {author} {\bibinfo {author} {\bibfnamefont {A.}~\bibnamefont
  {Laucht}} \emph {et~al.},\ }\bibfield  {title} {\enquote {\bibinfo {title}
  {Electrically controlling single-spin qubits in a continuous microwave
  field},}\ }\href {\doibase 10.1126/sciadv.1500022} {\bibfield  {journal}
  {\bibinfo  {journal} {Science Advances}\ }\textbf {\bibinfo {volume} {1}},\
  \bibinfo {pages} {e1500022} (\bibinfo {year} {2015})}\BibitemShut {NoStop}%
\bibitem [{\citenamefont {Ansaloni}\ \emph {et~al.}(2020)\citenamefont
  {Ansaloni} \emph {et~al.}}]{ansaloni2020natcomm}%
  \BibitemOpen
  \bibfield  {author} {\bibinfo {author} {\bibfnamefont {F.}~\bibnamefont
  {Ansaloni}} \emph {et~al.},\ }\bibfield  {title} {\enquote {\bibinfo {title}
  {Single-electron operations in a foundry-fabricated array of quantum dots},}\
  }\href {\doibase 10.1038/s41467-020-20280-3} {\bibfield  {journal} {\bibinfo
  {journal} {Nature Communications}\ }\textbf {\bibinfo {volume} {11}},\
  \bibinfo {pages} {6399} (\bibinfo {year} {2020})}\BibitemShut {NoStop}%
\bibitem [{\citenamefont {Shearrow}\ \emph {et~al.}(2018)\citenamefont
  {Shearrow} \emph {et~al.}}]{shearrow2018apl}%
  \BibitemOpen
  \bibfield  {author} {\bibinfo {author} {\bibfnamefont {A.}~\bibnamefont
  {Shearrow}} \emph {et~al.},\ }\bibfield  {title} {\enquote {\bibinfo {title}
  {Atomic layer deposition of titanium nitride for quantum circuits},}\ }\href
  {\doibase 10.1063/1.5053461} {\bibfield  {journal} {\bibinfo  {journal}
  {Applied Physics Letters}\ }\textbf {\bibinfo {volume} {113}},\ \bibinfo
  {pages} {212601} (\bibinfo {year} {2018})}\BibitemShut {NoStop}%
\bibitem [{\citenamefont {Schaal}\ \emph {et~al.}(2020)\citenamefont {Schaal}
  \emph {et~al.}}]{schaal2020prl}%
  \BibitemOpen
  \bibfield  {author} {\bibinfo {author} {\bibfnamefont {S.}~\bibnamefont
  {Schaal}} \emph {et~al.},\ }\bibfield  {title} {\enquote {\bibinfo {title}
  {{Fast Gate-Based Readout of Silicon Quantum Dots Using Josephson Parametric
  Amplification}},}\ }\href {\doibase 10.1103/PhysRevLett.124.067701}
  {\bibfield  {journal} {\bibinfo  {journal} {Phys. Rev. Lett.}\ }\textbf
  {\bibinfo {volume} {124}},\ \bibinfo {pages} {067701} (\bibinfo {year}
  {2020})}\BibitemShut {NoStop}%
\bibitem [{\citenamefont {{Beckers}}\ \emph {et~al.}(2018)\citenamefont
  {{Beckers}} \emph {et~al.}}]{beckers2018ted}%
  \BibitemOpen
  \bibfield  {author} {\bibinfo {author} {\bibfnamefont {A.}~\bibnamefont
  {{Beckers}}} \emph {et~al.},\ }\bibfield  {title} {\enquote {\bibinfo {title}
  {{Cryogenic MOS Transistor Model}},}\ }\href {\doibase
  10.1109/TED.2018.2854701} {\bibfield  {journal} {\bibinfo  {journal} {IEEE
  Transactions on Electron Devices}\ }\textbf {\bibinfo {volume} {65}},\
  \bibinfo {pages} {3617--3625} (\bibinfo {year} {2018})}\BibitemShut {NoStop}%
\bibitem [{\citenamefont {{Patra}}\ \emph {et~al.}(2020)\citenamefont {{Patra}}
  \emph {et~al.}}]{patra2020jeds}%
  \BibitemOpen
  \bibfield  {author} {\bibinfo {author} {\bibfnamefont {B.}~\bibnamefont
  {{Patra}}} \emph {et~al.},\ }\bibfield  {title} {\enquote {\bibinfo {title}
  {{Characterization and Analysis of On-Chip Microwave Passive Components at
  Cryogenic Temperatures}},}\ }\href {\doibase 10.1109/JEDS.2020.2986722}
  {\bibfield  {journal} {\bibinfo  {journal} {IEEE Journal of the Electron
  Devices Society}\ }\textbf {\bibinfo {volume} {8}},\ \bibinfo {pages}
  {448--456} (\bibinfo {year} {2020})}\BibitemShut {NoStop}%
\end{thebibliography}%

\clearpage

\section*{Methods}

\begin{center}
\textbf{Chip design and implementation}
\end{center}

The chip architecture was designed and simulated using Cadence Virtuoso, and industry-standard design tool for integrated circuit design and implementation. The initial design of quantum dot transistors, access transistors and passive elements, such as inductors, capacitors and resistors in the resonators and bias-tees, was based on existing device models at 300~K provided in the process design kit (PDK) of the 40-nm CMOS foundry. Subsequently, cryogenic models \cite{beckers2018ted,patra2020jeds} have been used to establish more predictive circuit design at 1-4~K temperatures, based on modified compact models for transistors, adjusted lumped-element equivalent models for resistors, capacitors and inductors, and electro-magnetic (EM) simulations with modified cryogenic substrate for custom microwave signal lines and connections. This allowed to provide a closer prediction of the chip performance at 50~mK, given that no established models exist for cryogenic circuit design at 1-4~K, let alone deep-cryogenic temperatures of 50~mK. Final simulations of the chip operation and performance have been performed using Cadence Virtuoso with such custom-modified cryogenic models and Keysight ADS Momentum (an industry-standard 3D planar EM solver) for electro-magnetic simulations of the whole chip, including all microwave lines, resonators, input ground-signal-ground (GSG) pads, bond-wires and external printed-circuit board (PCB) substrate, using modified material properties for cryogenic operation. The physical layout implementation was also performed with Cadence Virtuoso, while the physical design verification in terms of design rule check (DRC) and layout versus schematic (LVS) was performed using Mentor Graphics Calibre. The chip was finally fabricated by a standard VLSI manufacturing process in 40-nm technology in a multi-project wafer (MPW) without any custom modification.
The values of the circuit components in Fig.~\ref{fig:matrix}b are listed in Table~\ref{table:Spec}.

\vspace{\baselineskip}

\begin{center}
\textbf{Experimental setup}
\end{center}

\begin{figure}[H]
\centering
\includegraphics[width=0.45\columnwidth]{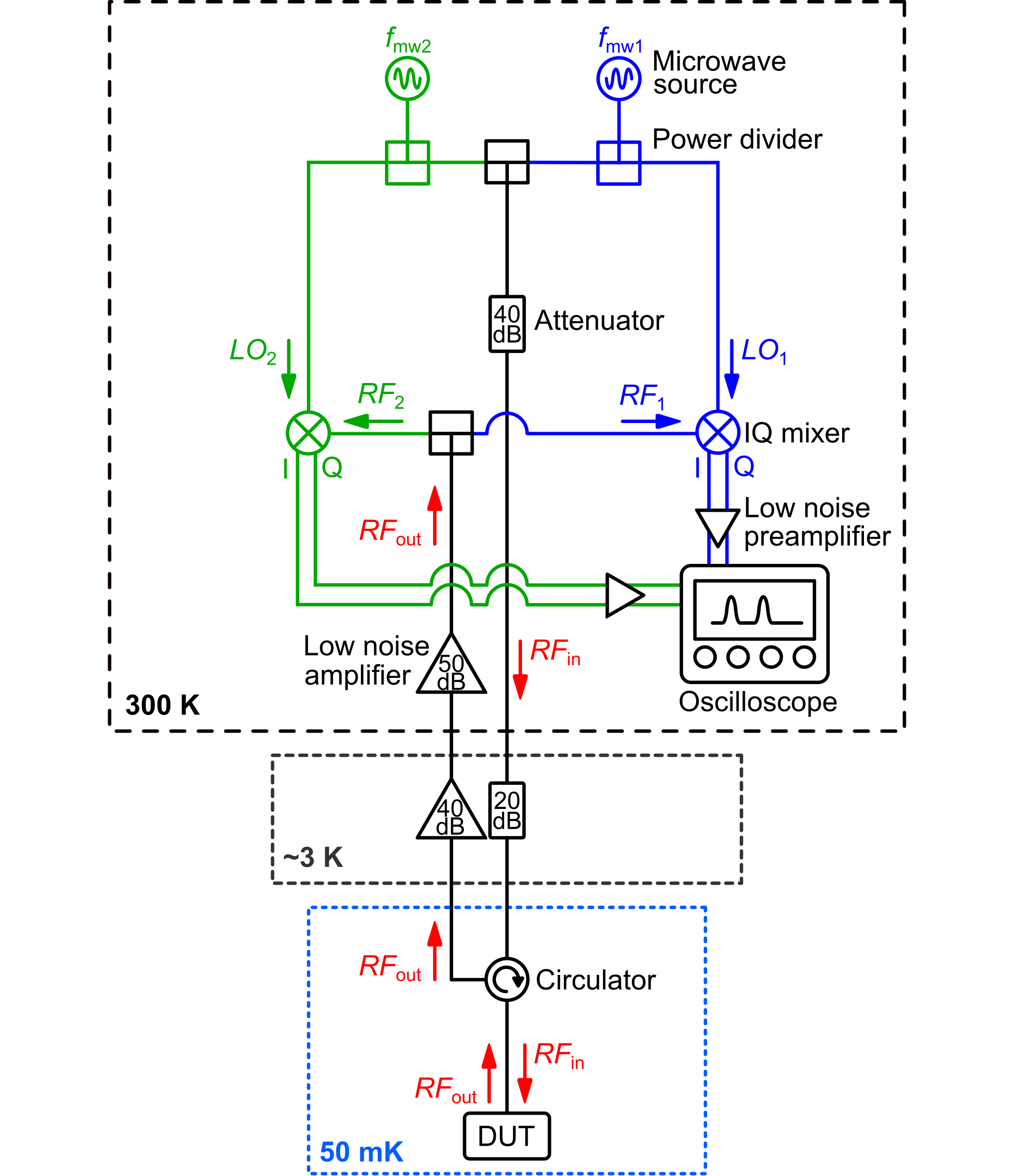}
\caption{\textbf{Experimental measurement setup.} Dilution fridge measurement set-up for the time- and frequency-multiplexed gate-based readout experiment. The DUT is placed at 50~mK on the sample stage of the dilution fridge. A low-noise amplifier and attenuators are placed at intermediate cryogenic temperatures ($\sim$ 3~K), and a cryogenic circulator is placed at the mixing chamber plate. At room temperature, two microwave signal sources are power-combined to generate two multiplexed single-tone probing signals $f_\mathrm{mw1}$ and $f_\mathrm{mw2}$, and two IQ mixers are used to demodulate the reflected signals $RF_\mathrm{1}$ and $RF_\mathrm{2}$ at the two frequencies, by using the microwaves sources as respective local oscillators $LO_\mathrm{1}$ and $LO_\mathrm{2}$. An oscilloscope acquires the I/Q outputs for each of the two tones.}
\label{fig:setup}
\end{figure}

All the measurements reported at 50~mK were performed in an Oxford Instruments Triton 200 dilution fridge. The designed chip was glued to a PCB and wire-bonded to a microwave transmission line on board. The device under test (DUT) was placed on the sample stage of the dilution fridge. All DC measurements were performed using DC lines in the fridge connected to a room temperature parameter analyzer HP~4156A, and/or source measure unit (SMU) Keithley~2400, used to apply and measure source, gate, drain voltages and currents. The gate DC voltages were alternatively applied using voltage sources HP 3245A and Keysight~33500B. Reflectometry measurements were performed with a modified setup, including a discrete cryogenic amplifier (Low Noise Factory LNF-LNC4\_8C) placed on a higher temperature stage (1-4~K), and a discrete cryogenic circulator (QUINSTAR QCY-G0400801) placed at the mixing chamber plate of the dilution fridge. A vector network analyzer (Rohde $\&$ Schwarz ZVA~24) has been used to measure the frequency spectrum. To perform gate-based reflectometry measurements, externally, additional low-noise amplifiers (PASTERNACK PE1524 and/or PE1522) and an IQ mixer (Marki IQ-0618MXP) have been used at room temperature to perform signal demodulation. Microwave sources (Anritsu~3692B and 3694C) were employed to provide the local oscillator signal for down-conversion in a homodyne scheme, and an oscilloscope (Teledyne LeCroy HDO4054A) was used to acquire the IF signals after they were amplified by low noise pre-amplifiers (Stanford Research Systems SR560). For time-multiplexing, arbitrary waveform generators (Keysight~33500B) were used to generate the square wave and ramp signal sequences used for $V_\mathrm{WLj}$ and $V_\mathrm{DLi}$, respectively. Finally, for (time- and) frequency-multiplexed measurements, two microwave signal sources have been used at room temperature with a power combiner (PASTERNACK~PE2068) to generate multiplexed single-tone microwave probing signals towards the DUT and two IQ mixers have been used to demodulate the two tones separately by using the same microwave sources as local oscillators, respectively. The obtained signals have been acquired by the oscilloscope simultaneously. The complete setup is shown in Fig.~\ref{fig:setup}.

\vspace{\baselineskip}

\begin{center}
\textbf{Data processing}
\end{center}

\begin{figure}[H]
\centering
\includegraphics[width=0.5\columnwidth]{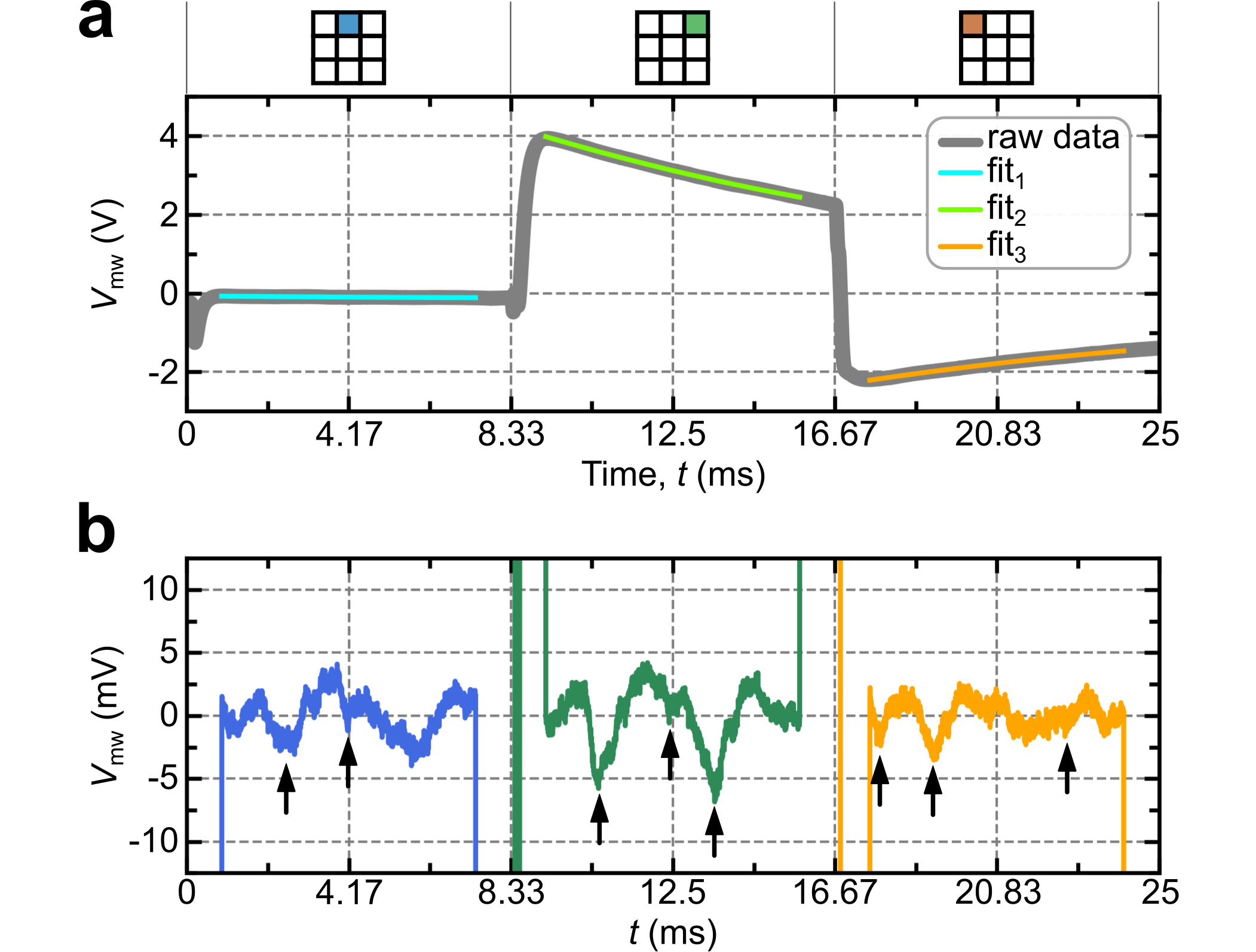}
\caption{\textbf{Data processing.} \textbf{a,} A single sweep of reflectometry signal $V_\mathrm{mw}$ (grey curve) as a function of time at $V_\mathrm{S}$=0~V corresponding to Fig.~\ref{fig:tmux}c. The data show the reflectometry signals from $Q_\mathrm{12}$ ($t$=0 to 8.33~ms), $Q_\mathrm{13}$ ($t$=8.33 to 16.67~ms), and $Q_\mathrm{11}$ ($t$=16.67 to 25 ms). The light blue, light green, and gold curves are fits to the data and used as backgrounds for data processing. \textbf{b,} Processed reflectometry signals of $Q_\mathrm{12}$ (blue), $Q_\mathrm{13}$ (green), and $Q_\mathrm{11}$ (gold) after subtracting the background fits. The arrows indicate Coulomb blockade peaks.}
\label{fig:dataprocessing}
\end{figure}

The time- (and frequency-)multiplexing readout data presented in this paper are processed data, and this is the data processing applied to it. In order to selectively address different columns in the quantum dot transistor array, square waves are applied to the word-lines when performing time-domain multiplexing readout. However, the rapid switches in the word-line voltages create large backgrounds which mask the reflectometry signals, as shown in Fig.~\ref{fig:dataprocessing}a. Hence, we use polynomial fits to the data as backgrounds. After subtracting the backgrounds from the original data, as shown in Fig.~\ref{fig:dataprocessing}b, the Coulomb blockade peaks are revealed and consistent with the individual reflectometry measurement for transistors $Q_\mathrm{12}$, $Q_\mathrm{13}$, and $Q_\mathrm{11}$ in Fig.~\ref{fig:tmux}a.

\vspace{\baselineskip}

\begin{center}
\textbf{Data analysis}
\end{center}

\begin{figure}[H]
\centering
\includegraphics[width=\textwidth]{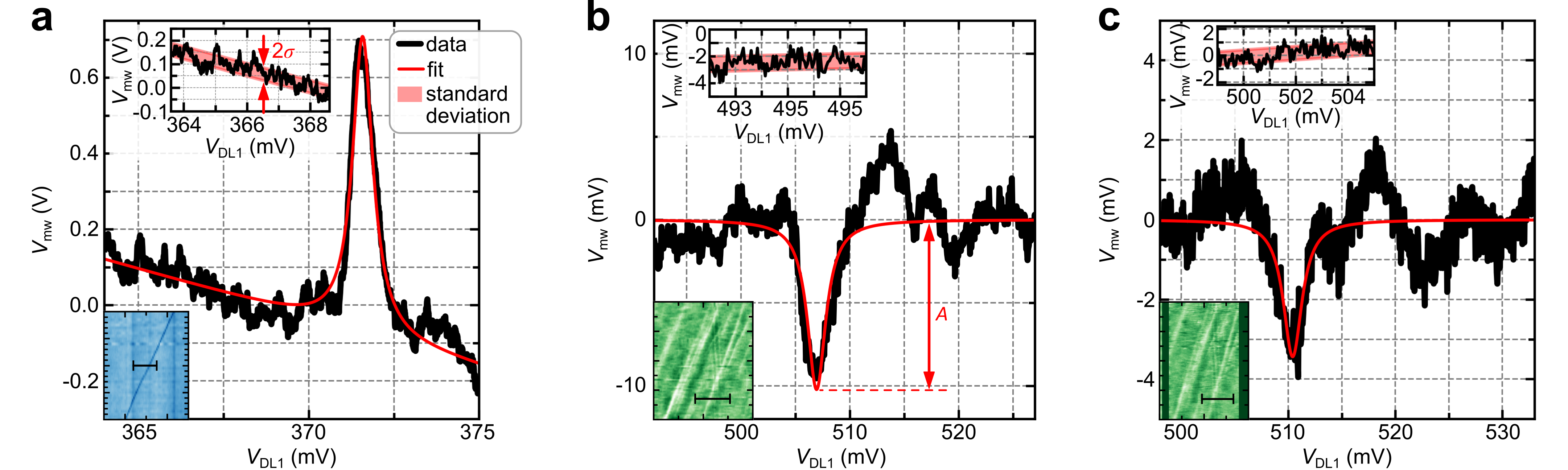}
\caption{\textbf{Signal-to-noise ratio analysis.} \textbf{a,} Trace of reflectometry signal $V_\mathrm{mw}$ as a function of $V_\mathrm{DL1}$ from an individual device measurement, corresponding to the black line section in the bottom left inset ($Q_\mathrm{13}$, left panel in Fig.~\ref{fig:reflectometry}c). The red curve is a Lorentzian fit to the experimental data. \textbf{b,} Trace of $V_\mathrm{mw}$ as a function of $V_\mathrm{DL1}$ for $Q_\mathrm{13}$ (middle panel in Fig.~\ref{fig:tmux}a) from an individual device measurement. $A$ is the Coulomb peak amplitude from the fit. \textbf{c,} Trace of $V_\mathrm{mw}$ as a function of $V_\mathrm{DL1}$ for $Q_\mathrm{13}$ in Fig.~\ref{fig:tmux}c from the time-domain multiplexing measurement. Top left inset: zoom in of the background noise. The measurement in \textbf{a} was performed in a different cool down thermal cycle from those in \textbf{b} and \textbf{c}, while the measurements in \textbf{b} and \textbf{c} are within the same cool down.}
\label{fig:SNR}
\end{figure}

The signal-to-noise ratio (SNR) can be derived from the reflectometry signal $V_\mathrm{mw}$, as shown in Fig.~\ref{fig:SNR}a, and SNR=\textit{A}/$\sigma$=28.7, where $A$ is the Coulomb blockade peak amplitude and $\sigma$ is the standard deviation of the noise signals. To extract the amplitude, we use a Lorentzian fit corresponding to the expected lineshape for lifetime broadened transitions~\cite{ahmed2018commphys}:

\begin{equation}
V_\mathrm{mw} \propto \frac{h\gamma}{(h\gamma)^2+(e\alpha(V_\mathrm{DL1}-V_\mathrm{DL1,offset}))^2},
\end{equation}

\noindent where $\gamma$ is the tunnel rate, $\alpha$ the lever arm or gate coupling factor and $V_\mathrm{DL1,offset}$ the data-line voltage at which the source and gate Fermi levels align. We compare the individual device measurement and the time-domain multiplexing measurement in Fig.~\ref{fig:SNR}b and Fig.~\ref{fig:SNR}c respectively, and show a benchmark in Table~\ref{table:SNR_benchmark}. Furthermore, we estimate the minimum integration time (defined as the time to reach SNR=1) from the bandwidth of the measurement and the number of averages. The tunnel rate for \textit{Source}-\textit{QD} is estimated as well. These results are shown in Table~\ref{table:SNR_benchmark}.

\vspace{\baselineskip}

\section*{Data availability}

The data that support the plots within this paper and all the findings of this study are available from the corresponding author upon reasonable request.

\section*{Acknowledgments}

We are grateful to Simon Schaal for providing useful comments.
\newline\indent
The research leading to these results has received funding from the European Union's Horizon 2020 research and innovation programme under grant agreement No. 688539 and 951852. M. F. G.-Z. acknowledges support from the Royal Society.

\section*{Author contributions}

A. R., M. F. G.-Z. and E. C. conceived the architecture and devised the experiments; A. R. and Y. P. designed the chip with inputs from E. C. and M. F. G.-Z.; T.-Y. Y., J. M., M. F. G.-Z. and A. R. performed the experiments and analyzed the results; A. R., T.-Y. Y. and M. F. G.-Z. wrote the manuscript with inputs from all the coauthors, M. F. G.-Z. and E. C. supervised all the experiments.

\section*{Competing interests}

The authors declare no competing interests.

\section*{Additional information}

\textbf{Supplementary information} is available for this paper.

\textbf{Correspondence} should be addressed to A. R. (andrea.ruffino@epfl.ch).

\textbf{Request for materials} should be addressed to T.-Y. Y. (tyy20@cam.ac.uk).

\clearpage

\section*{Supplementary Information}

\begin{center}
\textbf{S1. Retention time study}
\end{center}

\begin{figure}[H]
\centering
\includegraphics[width=\textwidth]{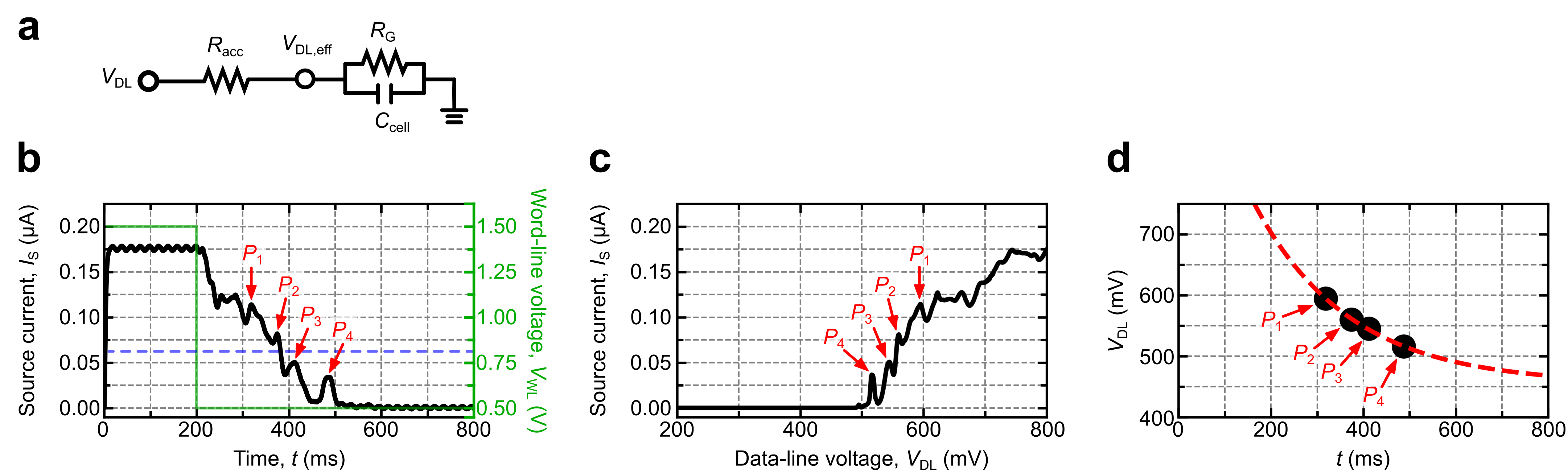}
\caption{\textbf{Retention time experiment.} \textbf{a,} Equivalent circuit of a single cell. \textbf{b,} Source current ($I_\mathrm{S}$) as a function of time ($t$) with the application of word-line voltage $V_\mathrm{WLj}^{High}$=1.49~V for $t$=0 to 200~ms and $V_\mathrm{WLj}^{Low}$=0.5~V afterwards (green line), and a constant data-line voltage $V_\mathrm{DL}$=0.8~V (blue dashed line). \textbf{c,} $I_\mathrm{S}$ as a function of $V_\mathrm{DL}$ at $V_\mathrm{WL}$=1.49~V. \textbf{d,} Locations of Coulomb blockade peaks $P_\mathrm{1}$, $P_\mathrm{2}$, $P_\mathrm{3}$, and $P_\mathrm{4}$ in time (in \textbf{b}), as a function of the corresponding Coulomb peaks in voltage (in \textbf{c}). The red dashed line is an exponential decay fit.}
\label{fig:Retention}
\end{figure}

We characterize the charge retention time for an individual cell at 50~mK. A simple equivalent circuit model is shown in Fig.~\ref{fig:Retention}a \cite{schaal2019natel}. To extract the retention time, we apply the sequence (i) charging and (ii) discharging. For (i) charging: the cell is firstly charged by applying $V_\mathrm{WL}$=1.49~V, much higher than the threshold voltage of the access transistor, while $V_\mathrm{DL}$=0.8~V, as in Fig.~\ref{fig:Retention}b, setting the access transistor well in the ON state. This is then followed by (ii) discharging: $V_\mathrm{WL}$ is reduced to 0.5~V, where the access transistor is highly resistive. The effective voltage on the QD transistor gate $V_\mathrm{G}$ as a function of time can be expressed as:

\begin{equation}
V_\mathrm{G}=V_\mathrm{0}\left[1+\frac{R_\mathrm{acc}}{R_\mathrm{G}}exp\left(-\frac{t}{\tau}\right)\right],
\label{equ:decay}
\end{equation}

\noindent where $V_\mathrm{0}=\frac{V_\mathrm{DL}R_\mathrm{G}}{R_\mathrm{acc}+R_\mathrm{G}}$ is the equilibrium voltage at the QD transistor gate at $t\rightarrow\infty$, and $R_\mathrm{G}$ and $R_\mathrm{acc}$ are the gate leakage resistance of the QD transistor and the channel resistance of the access transistor, respectively. $\tau=\frac{C_\mathrm{cell}R_\mathrm{G}R_\mathrm{acc}}{R_\mathrm{G}+R_\mathrm{acc}}$ is the circuit time constant, i.e., retention time, where $C_\mathrm{cell}$ is the parallel sum of the QD transistor gate capacitance and storage capacitance $C_\mathrm{C}$ in Fig.~\ref{fig:matrix}b. By monitoring $I_\mathrm{S}$ after $V_\mathrm{WL}$ is switched from 1.49~V to 0.5~V, Coulomb oscillations are observed as a function of time due to the decay of $V_\mathrm{G}$ in time, as shown in Fig.~\ref{fig:Retention}b. The observed Coulomb peaks in the time domain, marked as $P_\mathrm{1}$, $P_\mathrm{2}$, $P_\mathrm{3}$, and $P_\mathrm{4}$, have their counterparts in the voltage domain, as shown in Fig.~\ref{fig:Retention}c. Combining the marked Coulomb peaks in time and voltage domains, we fit the data points to Equation~\eqref{equ:decay} and find the time constant $\tau\approx$207~ms in this cell, as shown in Fig.~\ref{fig:Retention}d. From Fig.~\ref{fig:matrix}d in the main text, we deduce that at $V_\mathrm{WL}$=0.5~V, $R_\mathrm{acc}>R_\mathrm{G}$ and leakage occurs primarily through the gate resistance of the QD device. Considering $C_\mathrm{cell} \sim C_\mathrm{C}$=200~fF, we obtain $R_\mathrm{G}\sim$1 T$\Omega$. Further improvements in the gate voltage retention time will require increasing $R_\mathrm{G}$.
\newline\indent
It is also worth mentioning that limited charge injection is visible in Fig.~\ref{fig:Retention}b after switching. The design only uses a single nMOS pass transistor, instead of a transmission gate (nMOS and pMOS) for the access transistor, but additional dummy single-finger transistors have been included in the layout next to the access transistor to absorb and minimize the charge injected by switching.

\vspace{\baselineskip}

\begin{center}
\textbf{S2. Resonator characterization at room temperature}
\end{center}

\begin{figure}[H]
\centering
\includegraphics[width=0.5\columnwidth]{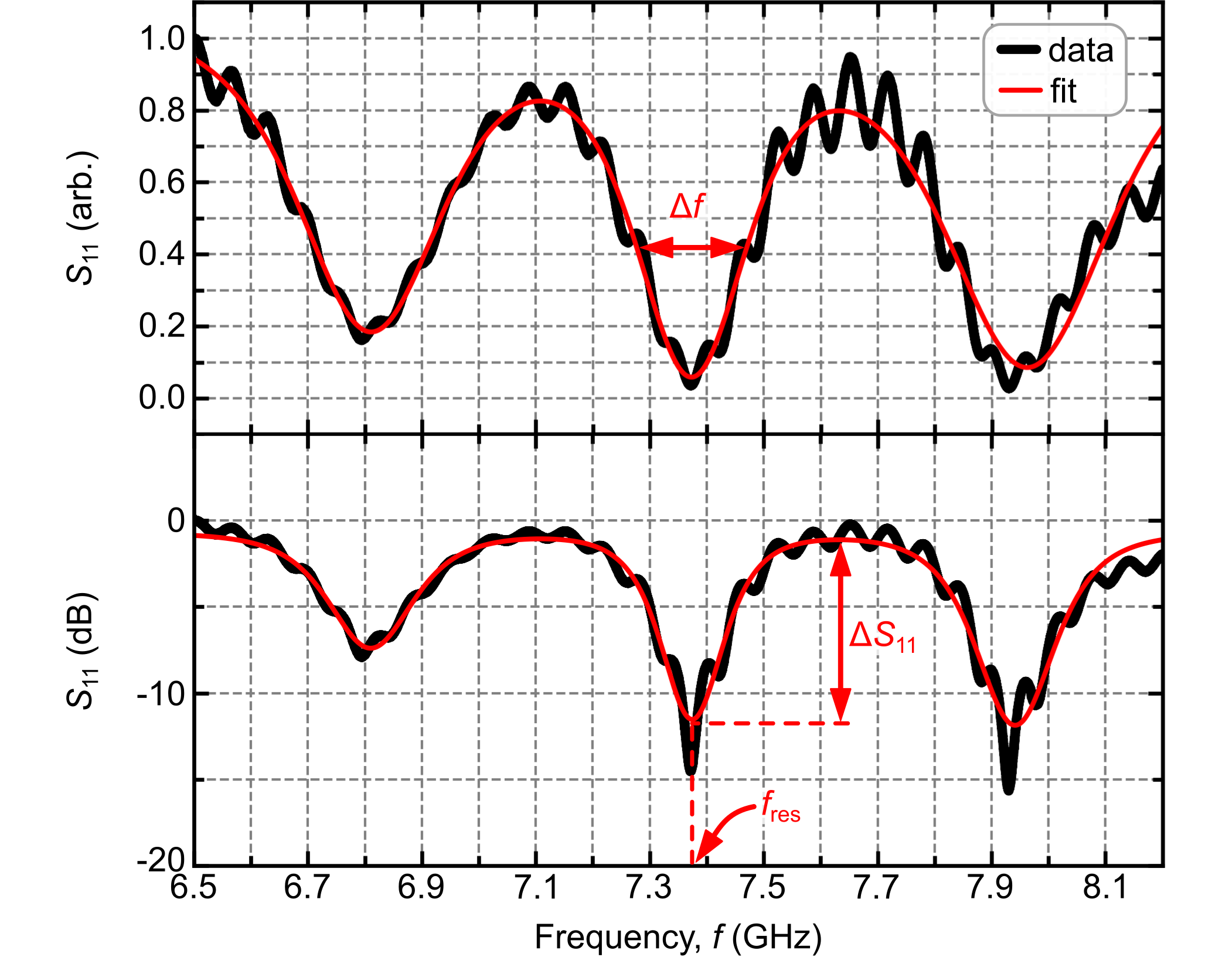}
\caption{\textbf{Frequency spectrum analysis.} Frequency spectrum of integrated LC resonators at 300~K. The black traces are the experimental data in linear scale (upper panel) and decibel scale (bottom panel), and the red curves are squared Lorentzian fits. $\Delta S_\mathrm{11}$, $f_\mathrm{res}$, and $\Delta f$ are the dip amplitude, center of the dip, and the dip full width at half maximum from the fit, respectively.}
\label{fig:frequency}
\end{figure}

We characterize the integrated LC resonators by measuring the frequency spectrum, as shown in Fig.~\ref{fig:frequency}. We perform the measurement at room temperature by directly connecting a VNA to the microwave port on the PCB (\textit{Set-up A}). We perform numerical fittings to the data by initially estimating the resonance frequencies ($f_\mathrm{res}$) where the reflection coefficient has local minima, i.e., around (6.8, 7.4, 7.9)~GHz for ($Resonator_\mathrm{1}$, $Resonator_\mathrm{2}$, $Resonator_\mathrm{3}$), respectively. The resonance frequencies and the reflection coefficients at the resonance frequencies ($\Delta S_\mathrm{11}$) are then extracted from the fitted results (red curves in Fig.~\ref{fig:frequency}). Furthermore the quality factors (\textit{Q}) can be derived from $Q=f_\mathrm{res}/\Delta f$, where $\Delta f$ is the full width at half maximum of the fit. The characteristics of the three integrated resonators are listed in Table~\ref{table:Resonator}.
\newline\indent
It is worth mentioning that the resonators on chip are, instead of more conventional L-shaped LC matching networks, $\pi$ CLC matching networks ($C_\mathrm{S}$, $L$, $C_\mathrm{P}$), since the former would impose a fixed \textit{matching network} quality factor determined by the ratio of source (50~$\Omega$) and load impedance (the gate of the QD device), while the latter introduces an additional degree of freedom, thus allowing to independently determine the \textit{matching network} quality factor. Therefore, all components in the matching network are functional, not due to parasitics. Finally, however, the measured quality factor is in any case determined by the \textit{component} quality factor, in this case mostly by inductors.

\clearpage

\onecolumngrid

\section*{Extended Data}

\clearpage

\begin{table}[H]
\centering
\caption{\textbf{Values of the circuit components.}}
\includegraphics[width=0.5\columnwidth]{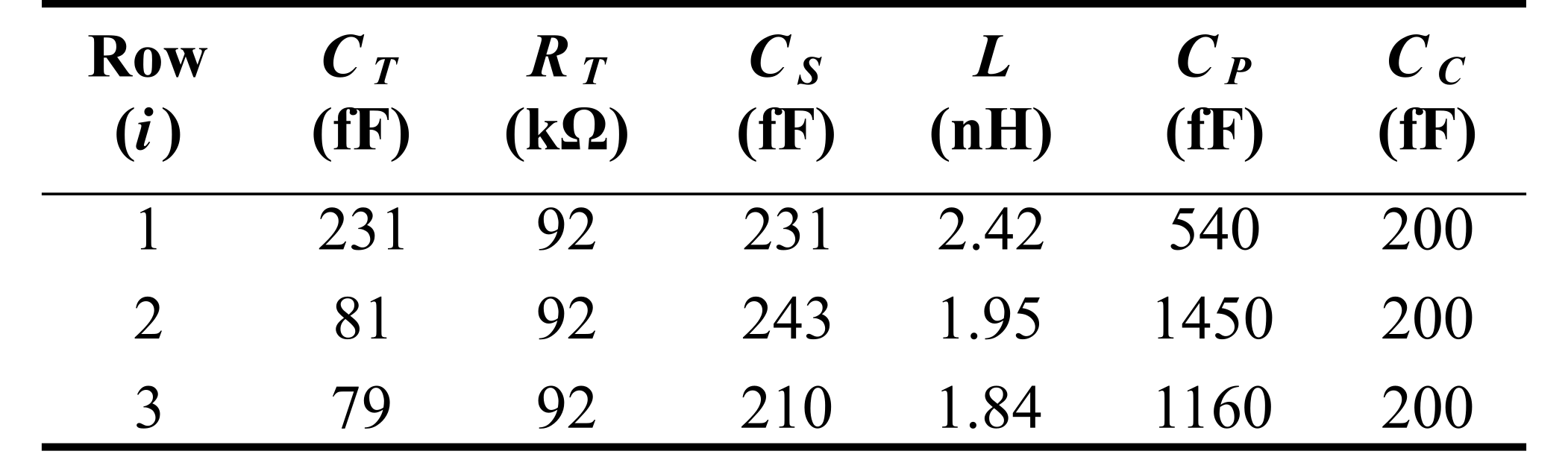}
\label{table:Spec}
\end{table}

\begin{table}[H]
\centering
\caption{\textbf{Benchmark of signal-to-noise ratio analysis for device $Q_\mathrm{13}$.} Coulomb peak amplitude (\textit{A}), peak full width at half maximum (FWHM), noise standard deviation $\sigma$, signal-to-noise ratio (SNR), measurement integration time ($t_\mathrm{int}$), estimated minimum integration time ($t_\mathrm{min}$), gate lever arm ($\alpha$), \textit{Source}-\textit{QD} electron tunnel rate, and coefficient of determination of the fit ($R^\mathrm{2}$) for multiple data sets.}
\includegraphics[width=\textwidth]{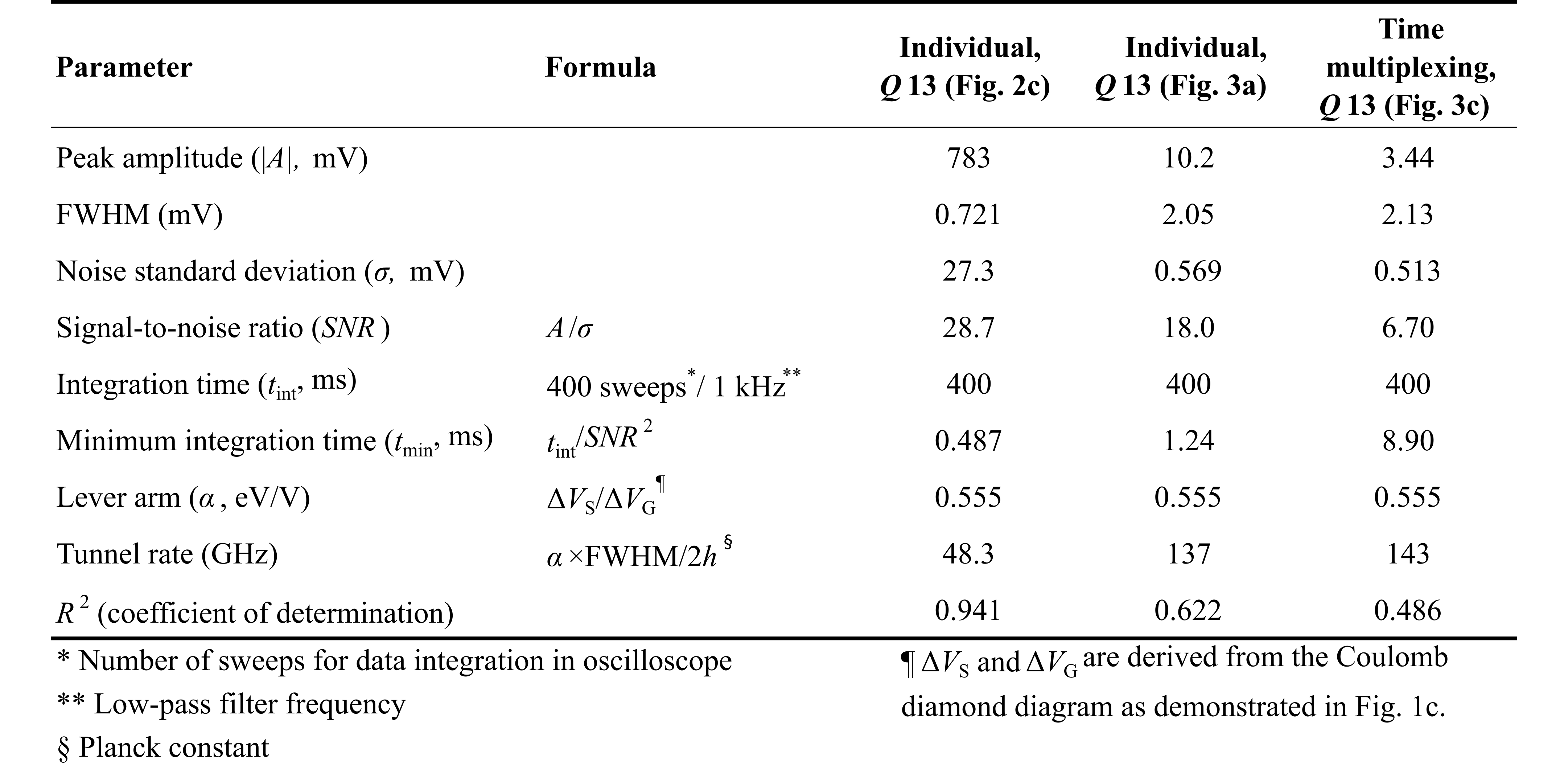}
\label{table:SNR_benchmark}
\end{table}

\begin{table}[H]
\centering
\caption{\textbf{Characteristics of the integrated LC resonators at 300 K.} Resonance frequency ($f_\mathrm{res}$), reflection coefficient at the resonance frequency ($\Delta S_\mathrm{11}$), and quality factor (\textit{Q}).}
\includegraphics[width=0.5\columnwidth]{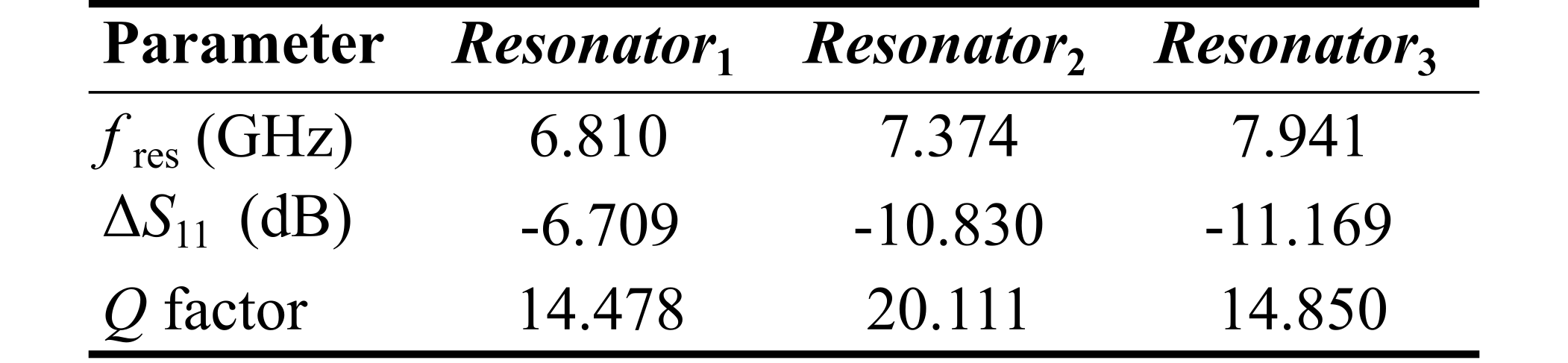}
\label{table:Resonator}
\end{table}

\end{document}